\documentclass[
reprint,
superscriptaddress,
nofootinbib,
amsmath,
amssymb,
aps,
floatfix,
]{revtex4-2}

\usepackage{graphicx}% Include figure files
\usepackage{dcolumn}% Align table columns on decimal point
\usepackage{bm}% bold math
\usepackage{epstopdf}
\usepackage{braket}
\usepackage{upgreek}

\usepackage{xcolor}
\usepackage{hyperref}
\hypersetup{colorlinks=True,allcolors=blue}

\begin{document}

\title{How to wire a 1000-qubit trapped ion quantum computer}
\author{M. Malinowski}
\email{mm@oxionics.com}
\affiliation{Oxford Ionics, Oxford, OX5 1PF, UK}
\author{D. T. C. Allcock}
\affiliation{Oxford Ionics, Oxford, OX5 1PF, UK}
\affiliation{Department of Physics, University of Oregon, Eugene, OR 97403}
\author{C. J. Ballance}
\affiliation{Oxford Ionics, Oxford, OX5 1PF, UK}
\affiliation{Department of Physics, University of Oxford, Clarendon Laboratory, Parks Road, Oxford, OX1 3PU, UK}

\begin{abstract}
One of the most formidable challenges of scaling up quantum computers is that of control signal delivery. Today’s small-scale quantum computers typically connect each qubit to one or more separate external signal sources. This approach is not scalable due to the I/O limitations of the qubit chip, necessitating the integration of control electronics. However, it is no small feat to shrink control electronics into a small package that is compatible with qubit chip fabrication and operation constraints without sacrificing performance. This so-called “wiring challenge” is likely to impact the development of more powerful quantum computers even in the near term.
In this paper, we address the wiring challenge of trapped-ion quantum computers. We describe a control architecture called WISE (Wiring using Integrated Switching Electronics), which significantly reduces the I/O requirements of ion trap quantum computing chips without compromising performance. Our method relies on judiciously integrating simple switching electronics into the ion trap chip – in a way that is compatible with its fabrication and operation constraints – while complex electronics remain external. To demonstrate its power, we describe how the WISE architecture can be used to operate a fully connected 1000-qubit trapped ion quantum computer using $\sim 200$ signal sources at a speed of $\sim 40-2600$ quantum gate layers per second.

\end{abstract}

\maketitle

\section{Introduction}
Trapped-ion qubits are one of the most promising approaches to quantum computing, especially in the NISQ era \cite{bhartiNoisyIntermediatescaleQuantum2022, bruzewiczTrappedIonQuantumComputing2019}. One of their main superpowers is ion transport, i.e. the ability to physically move ions in space \cite{homeCompleteMethodsSet2009}. This enables two powerful features:
\begin{enumerate}
    \item Qubit reconfiguration. Ion transport allows for flexible qubit routing -- that is, changing which qubit is coupled to which other qubits -- without relying on error-prone multi-qubit gates \cite{webberEfficientQubitRouting2020}. This allows for effective all-to-all connectivity even in a system composed of many uncoupled qubit registers.
    \item Transport-assisted gates. Ion transport allows for local control of quantum operation Rabi frequency even when the qubit drive operates at a fixed amplitude. This method can be employed with both laser and microwave qubit drives -- as long as they're spatially inhomogeneous -- and has been used for example in \cite{roweExperimentalViolationBell2001, declercqParallelTransportQuantum2016, seckSingleionAddressingTrap2020, tinkeyTransportEnabledEntanglingGate2022}.
\end{enumerate}

Qubit reconfiguration is the basis of the QCCD architecture \cite{winelandExperimentalIssuesCoherent1998, kielpinskiArchitectureLargescaleIontrap2002}, which is one of the most promising approaches to trapped-ion quantum computing. The effective all-to-all connectivity -- when combined with excellent coherence times \cite{wangSingleIonQubit2021, rusterLonglivedZeemanTrappedion2016, sepiolProbingQubitMemory2019}  -- is one of the reasons why trapped-ion systems achieve such high quantum volumes compared to other platforms \cite{pelofskeQuantumVolumePractice2022}. 
On the other hand, quantum gates in today’s QCCD systems are typically implemented by delivering localized externally modulated qubit drives (e.g. laser beams) to individual trap regions. Ion transport is only leveraged in a limited way, e.g. to move an ion away from a laser beam to switch off the interaction. However, as systems grow and off-chip local drive modulation becomes impractical, transport-assisted gates become a powerful tool, as they allow for local control with only a small number of global qubit drives. In other words, transport-assisted gates reduce the problem of implementing quantum gates at scale to the problem of ion transport at scale.

\begin{figure*}[!ht]
\centering
{\includegraphics[width=0.8 \textwidth]{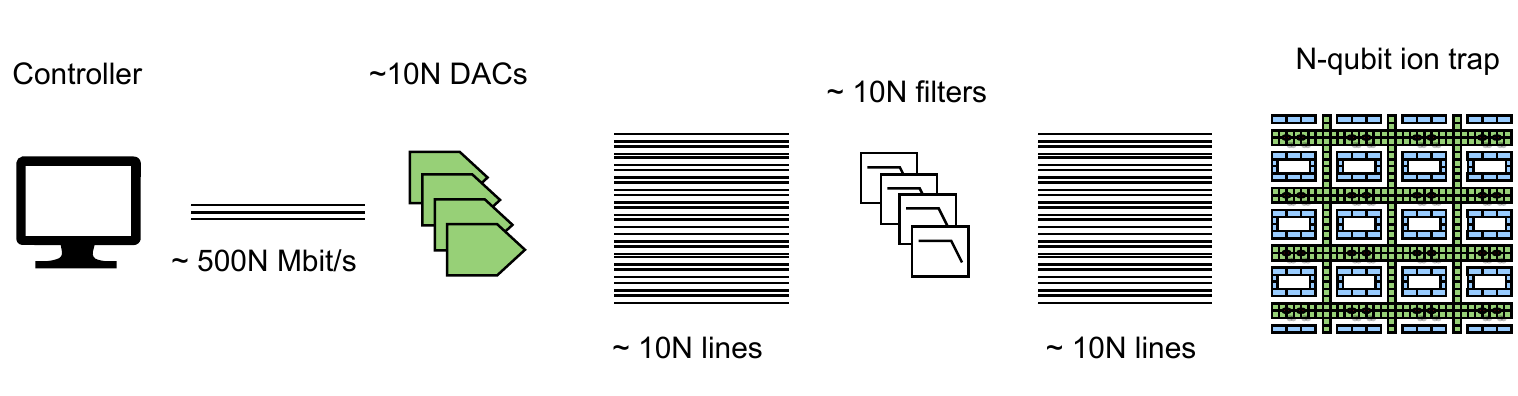}}
\caption{\label{fig:standard_approach} Illustration of the standard approach to the electrical wiring of trapped-ion quantum computers. Qubit transport in a $N$--qubit ion trap is achieved by using $\sim 10 N$ electrodes. Each electrode is wired through an individual filter to an individual DAC. The DAC output waveforms are set through a digital interface, with typical data flow rates $\sim 50$ Mbit/s per DAC. }
\end{figure*}

However, the ability to execute ion transport in large-scale systems is hindered by the wiring challenge. Fast, low-heating ion transport requires precise dynamical control of voltages on many electrodes \cite{aminiScalableIonTraps2010, sterkClosedloopOptimizationFast2022}. For example, QCCD architecture demonstrations from Pino \textit{et al.} \cite{pinoDemonstrationTrappedionQuantumCCD2021} and Kaushal \textit{et al.} \cite{kaushalShuttlingBasedTrappedIonQuantum2019} both used about 10 electrodes per qubit\footnote{For the sake of generality, we will use the term ``qubit transport" rather than ion transport. This accommodates for a possibility of encoding one qubit in a short ion chain, or that qubit ions may be always transported  in tandem with ancilla ions \cite{steaneHowBuild3002006}.}, each wired to a separate DAC outside of the vacuum system, see Fig.~\ref{fig:standard_approach}. This ``standard approach" is likely not scalable even to moderate system sizes. For example, a 1000-qubit chip would require approximately 10,000 input lines – more than today’s cutting-edge CPUs, and at a bleeding edge of packaging feasibility.
And while electrode co-wiring has been proposed \cite{alonsoQuantumControlMotional2013,holzphilipIonlatticeQuantumProcessors2019} and used \cite{revellePhoenixPeregrineIon2020, holz2DLinearTrap2020} to reduce the number of inputs per electrode, existing co-wiring approaches only allow for limited functionality, such as qubit storage of shuttling between remote zones.

One previously proposed solution to the wiring challenge is to form an integrated ``quantum processing unit" (QPU), which combines an ion trap chip with the DACs. The integration could be either monolithic \cite{stuartChipintegratedVoltageSources2019} or achieved by packaging together several independently fabricated chips \cite{guiseInVacuumActiveElectronics2014}. However, DAC integration comes with major challenges:
\begin{enumerate}
    \item  Power dissipation. For example, Stuart \textit{et al.} \cite{stuartChipintegratedVoltageSources2019} developed compact cryogenic (4K) DACs with a power consumption of $\sim 30$ mW per channel, or $\sim 300$ W for 1000 qubits. While this can be optimized, DAC power dissipation presents a significant challenge, especially in cryogenic environments, which are beneficial for reducing noise in trapped-ion systems.
    \item Data bandwidth. Fully flexible DAC control requires streaming large volumes of data to the QPU, typically $\sim 50$ Mbit/s per DAC \cite{kaushalShuttlingBasedTrappedIonQuantum2019, SinaraFastinoWiki2022}. For a 1000-qubit QPU, this corresponds to $\approx 500$ Gbit/s of data flow between the DAC and the control system. Designing an appropriate interface would not be trivial, and in practice might require integration of further digital electronics, for example, to store the waveforms \cite{bardinDesignCharacterization28nm2019}. This increases the complexity of QPU design and fabrication, and may limit the operation flexibility.
    \item Footprint. An integrated DAC will typically require a much larger chip area than the electrode itself. For example, a single unfiltered DAC block in \cite{stuartChipintegratedVoltageSources2019} used an area of $130 \ \upmu \text{m} \times 270 \ \upmu \text{m}$, and lower-noise DAC will require yet larger footprints, especially to accommodate integrated filters. Developing low-noise voltage sources with areas comparable to ion trap electrodes ($\sim 100 \ \upmu \text{m} \times 100 \ \upmu \text{m}$ or less) is thus a challenge in itself, and might require advanced techniques such as wafer stacking and trench capacitors \cite{allcockHeatingRateElectrode2012, romaszkoEngineeringMicrofabricatedIon2020, blainHybridMEMSCMOSIon2021}.
\end{enumerate}

Because of the problems highlighted above, current approaches are insufficient to address the wiring challenge, even for intermediate-scale QPUs with $\sim 1000$ qubits. 

In this paper, we present an architecture called WISE (Wiring using Integrated Switching Electronics), which addresses the challenge of wiring such intermediate-scale trapped-ion quantum computers. Our method relies on simple integrated electronics that entail minimal power dissipation, low data bandwidth, and small footprint, alleviating all the major challenges of DAC integration. In WISE, all complex high-footprint high-power electronics are placed off-chip, allowing for large-scale control without compromising performance.

The paper is structured as follows. In Sec.~\ref{sec:wise_method_overview}, we give an overview of the electrode wiring architecture and ion transport methods. We then describe how WISE can perform all the key operations of a large-scale QCCD architecture: arbitrary qubit reconfiguration (Sec.~\ref{sec:qubit_reconfiguration}) and transport-assisted gates (Sec.~\ref{sec:transport_assisted_gates}). Subsequently in Sec.~\ref{sec:implementation} we discuss the hardware implementation in more detail, demonstrating that WISE is indeed compatible with ion trap fabrication and operation constraints. Finally, we put it all together in Sec.~\ref{sec:wiring_1000_qubits}, where we describe how to build a 1000-qubit trapped-ion quantum computer, capable of running arbitrary quantum circuits with all-to-all connectivity but using only $\sim 200$ input lines.

\section{WISE method}
\label{sec:wise_method_overview}

\begin{figure*}[!ht]
\centering
{\includegraphics[width= \textwidth]{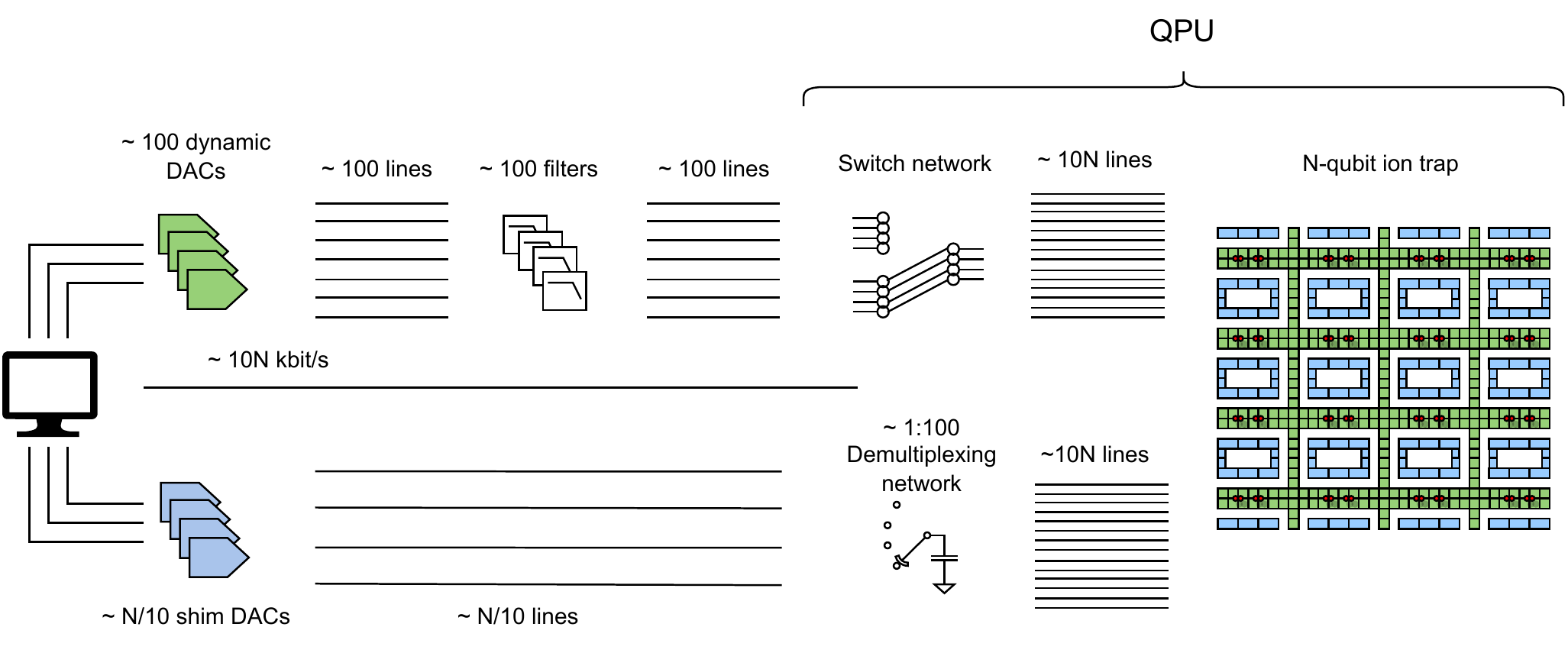}}
\caption{\label{fig:our_approach} High-level illustration of the WISE architecture. We form a Quantum Processing Unit (QPU) by combining an ion trap with a switch/demultiplexing network. In this way, all the high-density I/O is confined to the QPU, and the interface between the QPU and the outside world requires significantly fewer connections. We divide ion trap electrodes into dynamic ones (green, $\sim 10$ dynamic electrodes per qubits) and quasi-static ones, i.e. ``shims" (blue, $\sim 10$ shim electrodes per qubit). The dynamic electrodes are controlled by $\sim 100$ DACs, regardless of the system size, thanks to an integrated switch network. The number of shim DACs is reduced from one DAC per electrode in the standard approach to one DAC per $\sim 100$ electrodes in our approach, thanks to an integrated demultiplexing network. The high-speed digital interface is only required between the controllers and the DACs, while digital communication between the QPU and the controllers requires only small signal bandwidth, and can be done serially.}
\end{figure*}

In a QCCD device, ion positions are controlled by voltages applied to trap electrodes. These electrodes serve two different purposes. The first is \textit{dynamic}: to deliver time-varying waveforms which execute the desired transport primitives, such as shuttling \cite{waltherControllingFastTransport2012, bowler_coherent_2012}, merging, splitting \cite{kaufmannDynamicsControlFast2014, rusterExperimentalRealizationFast2014}, or crystal rotations \cite{kaufmannFastIonSwapping2017}. The second is \textit{quasi-static}: to compensate stray electric fields, generated for example by local charges and differences in work functions \cite{doretControllingTrappingPotentials2012, nadlingerMicromotionMinimisationSynchronous2021}. In WISE, each electrode is explicitly assigned to be either dynamic or quasi-static \cite{kienzlerQuantumHarmonicOscillator2015, decaroliMultiwaferIonTraps2021}. We then employ two techniques:
\begin{enumerate}
    \item \textit{Dynamic electrode parallelization}. Dynamic electrodes are co-wired to a fixed number of DACs, assigned through integrated switches.
    \item \textit{Quasi-static electrode demultiplexing}. Quasi-static ``shim" electrodes are controlled through a small number of DACs through integrated demultiplexers in a ``sample and hold'' fashion.
\end{enumerate}

In WISE, instead of DAC integration, we primarily use switch integration. Unlike DACs, switches require very small data input rates and can be operated with negligible power dissipation. Furthermore, they are relatively simple structures, made of a small number of transistors and inverters. Thus, they can be readily integrated into an ion-trap fabrication process~\cite{stuartChipintegratedVoltageSources2019}, and have been demonstrated to be cryogenically compatible, also in the context of quantum computing \cite{alonsoGenerationLargeCoherent2016, xuOnchipIntegrationSi2020}. The resulting high-level wiring architecture is shown in Fig.~\ref{fig:our_approach}.

\subsection{Dynamic electrode parallelization}

The main observation behind dynamic electrode parallelization is that the same transport operation can be executed in multiple areas of the chip at the same time by applying the same voltage waveforms to multiple electrodes. For example, parallel splits can be executed by connecting one set of DACs, delivering one set of split waveforms, to all the zones where we want to implement a split. Thus, instead of using one DAC per electrode, it suffices to use one DAC per waveform. Furthermore -- as we argue in Sec.~\ref{sec:reconfig_timing} -- while it might be optimal to execute different transport operations in different zones at the same time (for example, a split in some zones, and a merge in some other zones), it is nonetheless efficient to only perform only one transport operation at any given time (for example, a split in some zones, followed by a merge in some other zones). Thanks to these two ideas, a modest number of DACs suffices to execute arbitrary transport sequences.

Fig.~\ref{fig:our_approach} (top) illustrates the dynamic electrode wiring that leverages those insights. Outside the QPU, a fixed number of DACs are used to output qubit transport waveforms, while in the QPU, switches are used to select which electrode connects to which DAC. In subsequent sections, we show how dynamic electrode parallelization allows for efficient qubit reconfiguration in 1D and 2D ion traps (Sec.~\ref{sec:qubit_reconfiguration}), and why $\sim 100$ DACs suffice for the purpose, regardless of qubit number $N$. In Sec.~\ref{sec:transport_gates_parallel_dynamical} we discuss how dynamic electrode parallelization can be used to perform transport-assisted quantum gates. Finally, Sec.~\ref{sec:implementation_dynamic_parallelization} describes the hardware implementation of the dynamic electrode switch network.

\subsection{Quasi-static electrode demultiplexing}

In WISE, all electrodes which are not dynamic are quasi-static, meaning they are held at a constant voltage during any given qubit reconfiguration and during any given quantum gate \footnote{We say ``quasi" to indicate that the voltage can be adjusted between steps, for example, a different voltage set can be used for transport and for gates.}. The main insight is that holding a fixed voltage on electrodes can be less resource-intensive than applying time-dependent waveforms, because electrodes can maintain their set voltage even while disconnected from DACs. Thus, a single DAC can be demultiplexed to control multiple electrodes. In this mode of operation, the DAC output is connected to shim electrodes one by one, charging them to the necessary voltage before disconnecting. While the DAC is disconnected, the electrode is connected to ground through an integrated capacitor, which holds the DC voltage while acting as an RF shunt \cite{stuartChipintegratedVoltageSources2019}.

Since quantum gates and transport operations are very sensitive to electric field noise \cite{brownnuttIontrapMeasurementsElectricfield2015, talukdarImplicationsSurfaceNoise2016}, multiplexer operation can lead to additional errors, especially if the multiplexing frequency or the switching frequency is near the motional frequency of the ions ($\sim 1-10$ MHz). To alleviate that source of noise, we operate in a ``sample and hold" fashion, where all the electrodes are charged first, and the sensitive operations are performed with the multiplexer turned off.

Fig.~\ref{fig:our_approach} (bottom) illustrates the shim electrode wiring in the WISE architecture. Outside the QPU, a small number of DACs cycle through shim voltage setpoints, while in the QPU, demultiplexers are used to connect the DACs to on-chip capacitors and electrodes. In subsequent sections, we describe the role of shim electrodes in qubit reconfiguration (Sec.~\ref{sec:qubit_reconfiguration_shims}) and in transport-assisted gates Sec.~\ref{sec:transport_gates_shim_demux}. Finally, Sec.~\ref{sec:implementation_shim_demultiplexing} describes the hardware implementation of the shim demultiplexing network, and motivates the choice of $\sim 100$ shim electrodes per DAC.

\section{Qubit reconfiguration}
\label{sec:qubit_reconfiguration}

In this section, we show how dynamic electrode parallelization can be used for arbitrary and efficient reconfiguration in the QCCD architecture.
This is structured as follows. First, we describe how to perform a ``switchable swap" of two qubits in a linear (1D) array. Second, we show how to use switchable swaps to perform arbitrary reconfiguration of $N$ qubits in a linear array. Third, we extend the method to construct arbitrary reconfiguration of $N = m \times n$ qubits in a regular 2D array, where every qubit can be swapped with any of its four neighbors (i.e. one qubit per junction). Fourth, we extend the construction to a realistic $N$-qubit QCCD device, consisting of 2-qubit chains held in a 2D array with $k$ qubits per junction. Finally, we argue that the method is practical by calculating the expected runtime and error rate of a worst-case reconfiguration.

\subsection{Switchable swap}
\label{sec:switchable_swap}
A switchable swap refers to a procedure where two neighboring qubits are physically swapped conditioned on the settings of on-chip switches. 
\begin{figure}[!ht]
\centering
{\includegraphics[width=0.45 \textwidth]{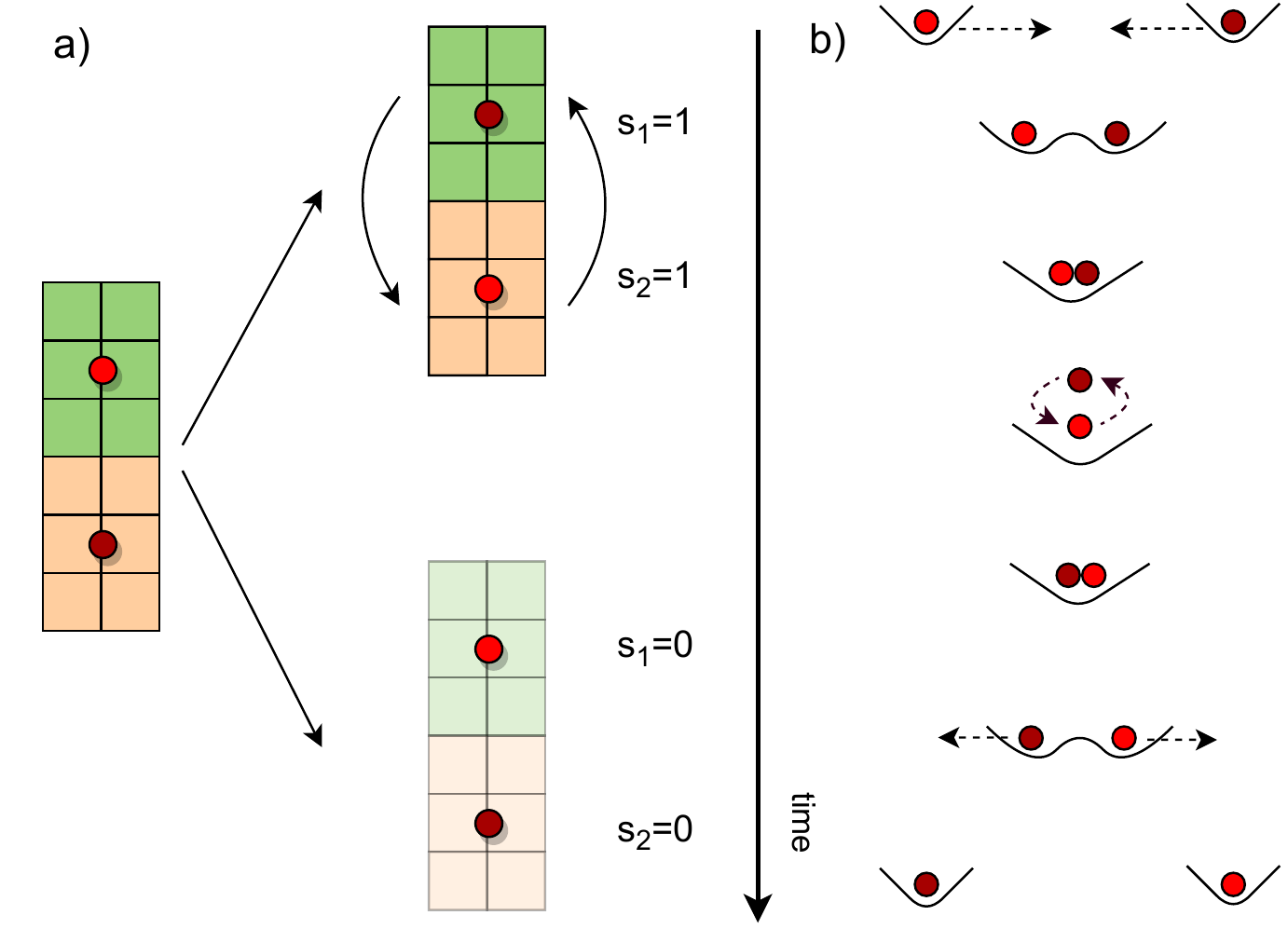}}
\caption{\label{fig:switchable_swap} a) Basic switchable swaps. Two qubits (red) are placed in neighboring zones (1,2), shown in green and orange, respectively. When switches $s_1,s_2$ are set to $(s_1, s_2) = (1,1)$, the zones are ``active", and the qubits undergo a swap. If however $(s_1, s_2) = (0,0)$, the zones are ``inactive" (drawn as partially transparent), and the qubits remain in their original locations. b) 1D swap sequence. Two qubits (red) in neighboring zones are swapped by bringing them together (shuttling) and merging their potential wells, followed by a crystal rotation, well split, and a shuttle step. At each step, the potential experienced by each ion is the net potential of all neighboring dynamic, shim, and RF electrodes.}
\end{figure}

Consider two qubits (1,2) held in separate zones (1,2), as illustrated in Fig.~\ref{fig:switchable_swap} a).  Each zone contains $N_{de/z} \sim 10$ dynamic electrodes. We implement a switchable swap as follows. If switches $(s_1,s_2)$ are set to $(1,1)$, we connect the dynamic electrodes in both zones to $2 \times N_{de/z} \sim 20$ DACs which play waveforms that result in a ``swap sequence'', such as shown in Fig.~\ref{fig:switchable_swap} b), which reorders the qubits. On the other hand, if switches are set to $(0,0)$, we connect the electrodes to $2 \times N_{de/z} \sim 20$ DACs which execute a ``stay still" sequence, keeping the qubits in place. This allows implementing a switchable swap using $4 \times N_{de/z} \sim 40$ DACs\footnote{In this particular example, the number of DACs could be reduced by half by simply leaving the electrodes unconnected whenever we want to execute a ``stay still" sequence. Further reduction in DAC could be achieved by wiring the odd and even zones to the same DACs, but with inverted mapping. However, we ignore these tricks in the rest of the paper for the sake of generality.} and 2 bits of information\footnote{Technically, as $(0,1)$ and $(1,0)$ are not valid control words, it suffices to send one bit here. However, controlling every switch individually will become important later, so we assume setting one switch requires one bit of information throughout the paper}.

\subsection{1D reconfiguration}
\label{sec:1d_reconfiguration}

\begin{figure}[!ht]
\centering
{\includegraphics[width=0.45 \textwidth]{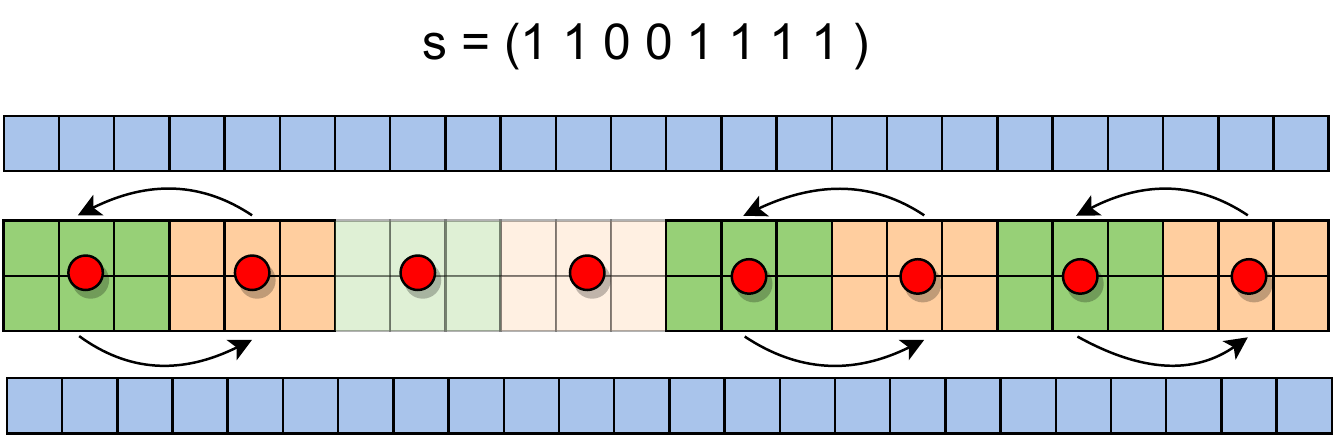}}
\caption{\label{fig:basic_swap_1d} Odd swap in a linear 1D array of $N=8$ qubits. In this example, the bit select word is $s = (1,1,0,0,1,1,1,1)$. Thus, zones 1,2,5,6,7,8 are active (their qubits undergo swaps), while zones 3,4 remain inactive (their qubits remain stationary). In addition to dynamic electrodes (shown in green and orange for odd and even zones respectively), RF electrodes are shown in white, and shim electrodes in blue.}
\end{figure}

We now enlarge the processor to be a linear repeating array with zones $i=1,2,\ldots, N$, as illustrated in Fig.~\ref{fig:basic_swap_1d}. Every second zone is connected to the same fixed set of DACs, i.e. if $s_{i} = s_{i+2}$, then every dynamic electrode in zone $i+2$ executes the same waveform as the corresponding dynamic electrode in zone $i$. Thus, by playing the same waveforms as in Sec.~\ref{sec:switchable_swap}, and sending an $N$-bit word $s = (s_1, s_2, \ldots, s_N)$ to the QPU, we can implement a switchable swap of qubits $(i, i+1)$ for every odd $i$ in parallel. We call this step an ``odd swap". Similarly, an ``even swap" - a switchable swap of qubits $(i, i+1)$ for every even $i$ in parallel - can be accomplished by playing a different ``swap waveform" and sending another $N$-bit word $s$. Implementing an arbitrary odd or even swap in a $N$-qubit array requires $4\times N_{de/z} \sim 40$ DACs, the same as a switchable swap, and $N$ bits of information.

As is well known, these odd/even swap primitives suffice to perform arbitrary qubit reconfiguration in 1D using an algorithm known as ``odd-even sort" \cite{WikipediaOddevenSort2022}. Specifically, denote the current qubit configuration as $\vec{x} = (x_1,x_2,\ldots,x_N)$, and the target configuration as $\pi(\vec{x}) = (\pi(x_1),\pi(x_2),\ldots,\pi(x_N))$. In the first time step, qubits $x_i$ and $x_{i+1}$ swapped if $\pi(x_i) > \pi(x_{i+1})$ for every odd $i$. In the second time step step, qubits $x_i$ and $x_{i+1}$ swapped if $\pi(x_i) > \pi(x_{i+1})$ for every even $i$. These steps are repeated until $\vec{x} = \pi(\vec{x})$. The odd-even sort is the time-optimal method of sorting qubits in 1D, with a worst-case run time of $N$ time-steps.

\subsection{2D regular array reconfiguration}
\label{sec:2d_regular_array_reconfig}
\begin{figure}[!ht]
\centering
{\includegraphics[width=0.45 \textwidth]{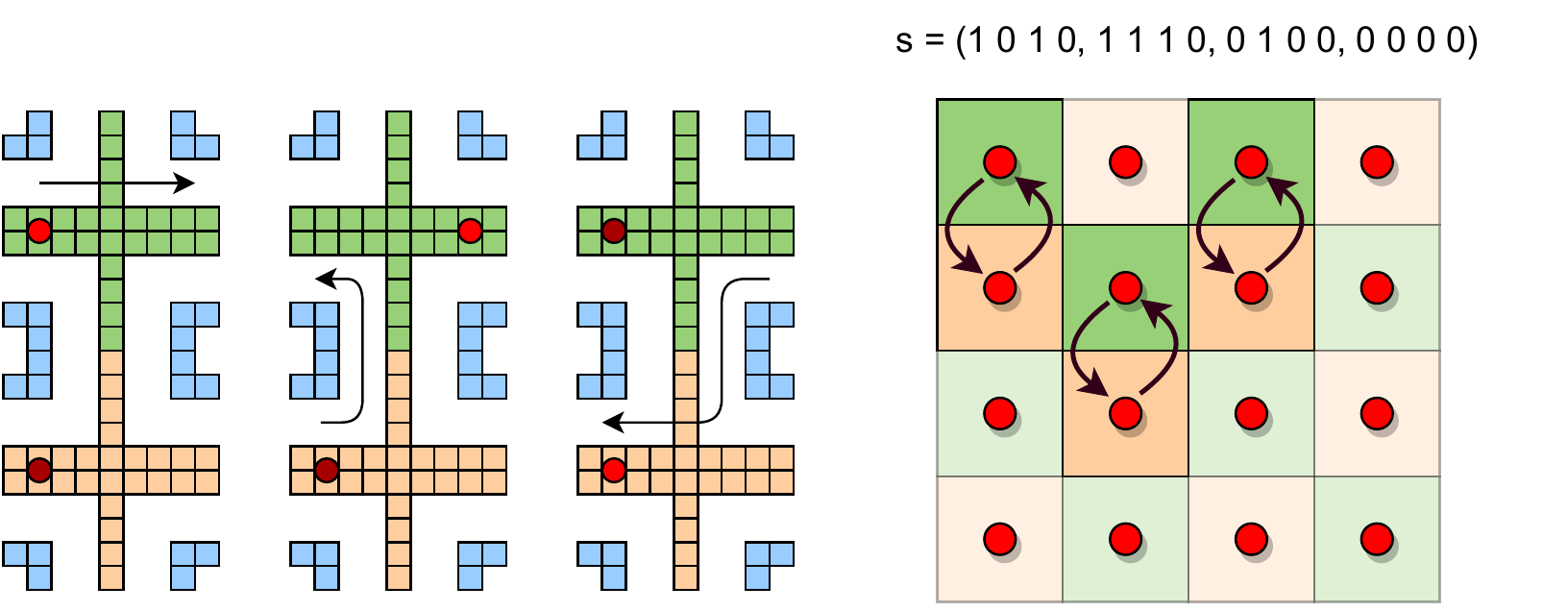}}
\caption{\label{fig:2d_array_swaps} a) 2D qubit swap sequence. Two qubits in neighboring junctions can be swapped without crystal rotations as a sequence of three shuttling steps. b) 2D qubit swap.
During a 2D swap, zone $(i,j)$ is active (solid color) iff $s_{i,j} = 1$, in which case its qubit undergoes a swap operation. In the image above, the ion trap is a regular 4x4 array, and the bit-select word $s = (s_{1,1}, s_{1,2}, \ldots, s_{2,1} \ldots, s_{4,4})$ has length 16. This particular swap step is an odd vertical swap, i.e. zones $(i,j)$ and $(i,j+1)$ undergo a swap iff $i+j$ is even and $(s_{i,j}, s_{i,j+1}) = (1,1)$.}
\end{figure}

Consider a rectangular 2D grid of qubits, where every qubit can be swapped with any of its four neighbors. This corresponds to an ion trap tiled with X-junctions, with one qubit per junction \cite{webberEfficientQubitRouting2020}. In this case, qubit swap can still be performed using a swap sequence as in Fig.~\ref{fig:switchable_swap} b). Alternatively, it can be executed as a sequence of shuttling steps as illustrated in Fig.~\ref{fig:2d_array_swaps} a). Regardless of the physical implementation, we can still consider a switchable qubit swap to be a logical primitive, and the swap can be either horizontal (in a row) or vertical (in a column).

The 1D reconfiguration algorithm can be extended onto a regular 2D grid as follows \cite{alonRoutingPermutationsGraphs1994, childsCircuitTransformationsQuantum2019, banerjeeLocalityawareQubitRouting2022}. Consider a grid of $m \times n$ qubits arranged in a 2D array. We enumerate zones as $(i,j)$, where $i = 1, 2, \ldots, m$ and $j = 1, 2, \ldots, n$. A zone is considered ``odd" if $i+j$ is odd, and ``even'' otherwise, as shown in Fig.~\ref{fig:2d_array_swaps} b). We achieve arbitrary qubit reconfiguration as follows. First, we rearrange the qubits in every row in parallel such that, for every column, the target row of every qubit is different, which is always possible thanks to Hall's marriage theorem \cite{banerjeeLocalityawareQubitRouting2022}. This rearrangement proceeds by horizontal odd-even swap as outlined in Sec.~\ref{sec:1d_reconfiguration}, and thus takes at most $m$ time-steps. Afterward, we rearrange the qubits in every column in parallel to place every qubit in the target row. This is executed by vertical odd-even swap and takes at most $n$ time steps. Finally, we proceed with the final row-wise rearrangement, which takes at most $n$ time steps. Thus, the WISE architecture allows for arbitrary permutation of $N = m \times n$ qubits in 2D in at most $2m+n$ time steps\footnote{Changing the permutation order to column-row-column allows for arbitrary permutation in at most $2n+m$ steps, which is more efficient if $m > n$} and two types of zones.

\subsection{Realistic 2D array reconfiguration}
\label{sec:realistic_2d_reconfig}

The regular 2D array described in Sec.~\ref{sec:2d_regular_array_reconfig} is not a preferred arrangement of zones for several reasons. First, junctions typically require a larger electrode count and footprint than linear segments. Thus, allocating one junction per qubit may be exceedingly costly, and it will likely be preferable to operate with $k>1$ qubits per junction. Second, quantum computing requires not just an array of individual qubits, but an array of qubit chains to facilitate multi-qubit gates. Thus, any reconfiguration method must consider how qubits come together in larger chains. Finally, a QPU might require specialized zones, such as an ``ion loading zones" \cite{shiAblationLoadingBarium2023} or ``qubit readout zones" \cite{ghadimiScalableIonPhoton2017}. A practical reconfiguration method should thus allow for different zone types.

\begin{figure}[!ht]
\centering
{\includegraphics[width = 0.45 \textwidth]{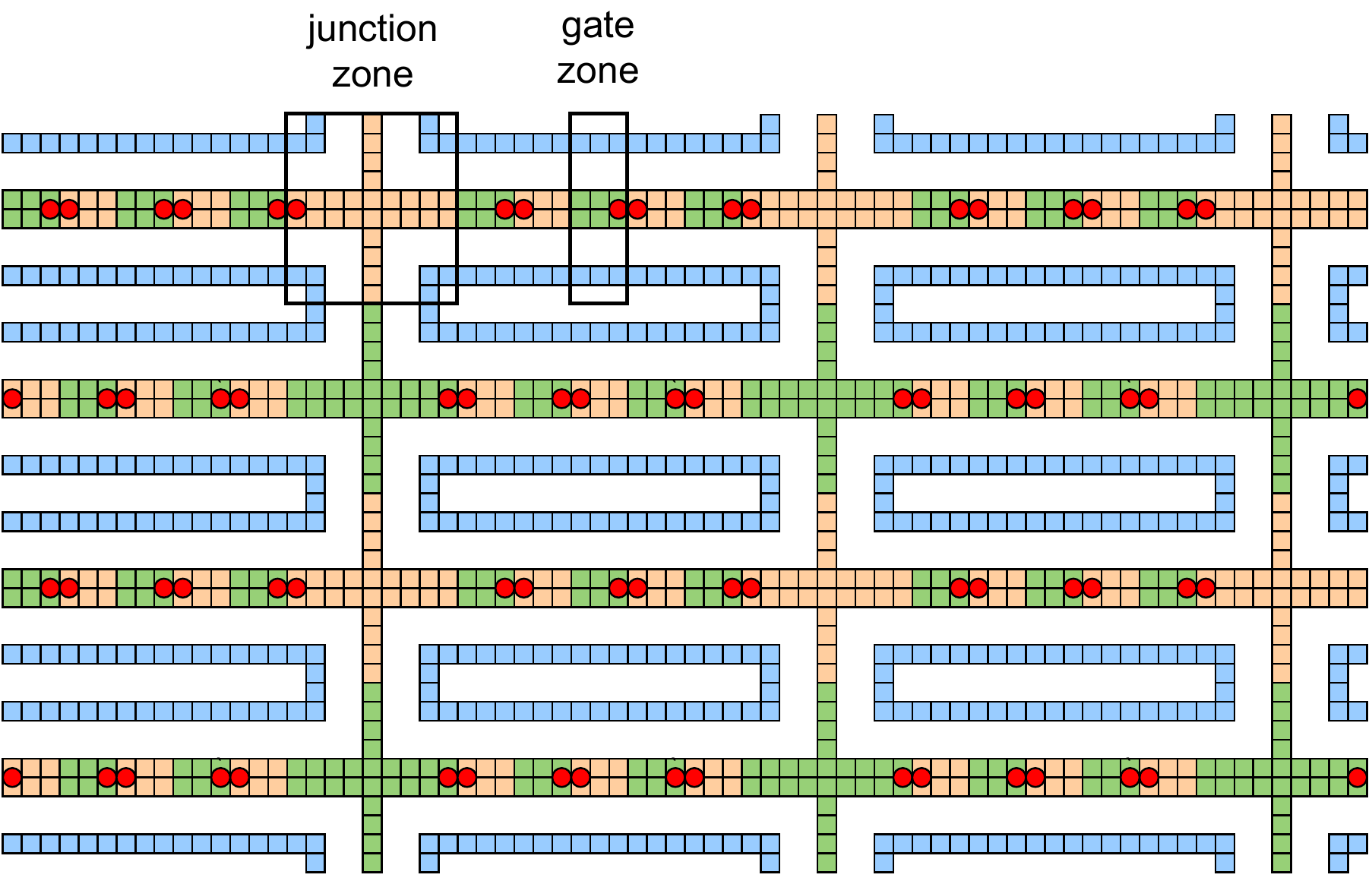}}
\caption{\label{fig:realistic_2d_trap} Schematic illustration of a 2D ion trap with $k=6$ qubits per junction. We distinguish four zone types: junction odd (orange), junction even (green), gate odd (green), and gate even (orange). Before and after reconfigurations, qubits are stored in the linear segments in small chains to allow for multi-qubit quantum gates.}
\end{figure}

Fortunately, the methods presented above can be easily extended to such more realistic traps. As an example, consider a 2D ion trap with $k=6$ qubits per junction, configured to perform quantum gates on two-qubit chains stored in ``gate zones". An illustration of the trap and qubits is shown in Fig.~\ref{fig:realistic_2d_trap}.

\begin{figure*}[!ht]
\centering
{\includegraphics[width = 0.9 \textwidth]{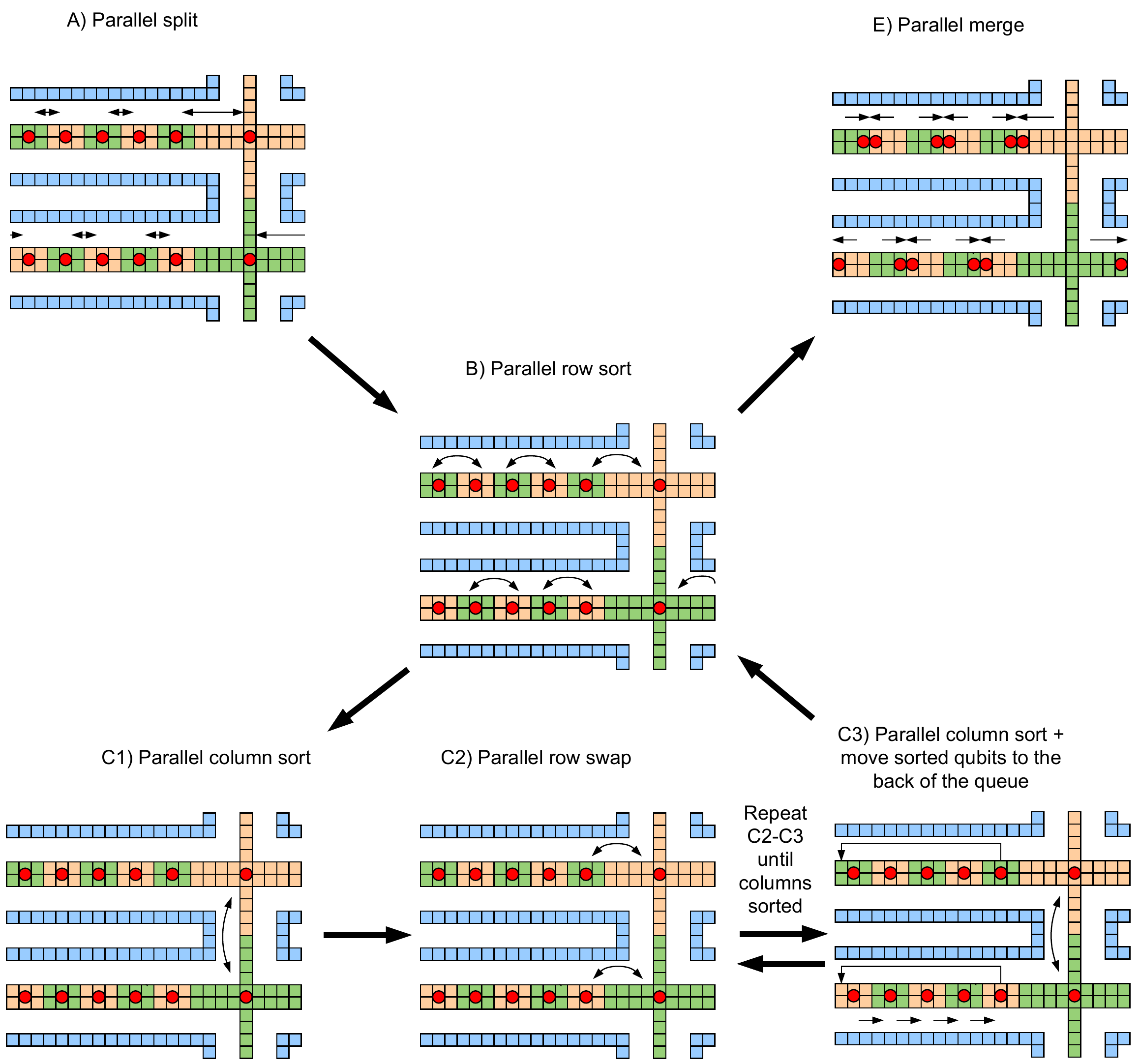}}
\caption{\label{fig:realistic_2d_trap_sort} Sort algorithm for a realistic 2D ion trap with $k=6$ qubits per junction. A) First, two-qubit chains in linear segments are split in parallel such as to place every qubit in a different zone. B) Parallel row-wise odd/even sort is used exactly as if the array were regular, rearranging the qubits such that qubits in the same column have unique destination rows. This takes at most $m$ time steps. C) Parallel column odd/even sort is applied on every $k$th qubit in sequence. This takes $k \times n$ time-steps \footnote{Technically, additional $k$ time-steps are needed for moving qubits in- and out- of junctions (step C2 in Fig.~\ref{fig:realistic_2d_trap_sort}). Furthermore, if $k$ is large enough, moving sorted qubits to the back of the queue (step C3 in Fig.~\ref{fig:realistic_2d_trap_sort}) may occasionally hold up the parallel column sort. On the other hand, the algorithm can be slightly sped up by other means, such as initiating the sort of the next column while the previous one is not fully sorted. We leave these caveats for future work.}. D) Parallel column odd/even sort is used exactly as if the array were regular. E) Finally, qubits are re-merged into two-qubit chains between gate zones in parallel.}
\end{figure*}

In order to perform arbitrary reconfiguration using a dynamic electrode parallelization, we divide the trap into four zone types -- junction odd, junction even, gate odd, gate even -- arranged as shown in Fig.~\ref{fig:realistic_2d_trap}. Since every odd zone only neighbors even zones, row-wise and column-wise odd-even sort operations remain unchanged.  We assume each gate zone contains $N_{de/gz} \sim 10$ dynamic electrodes, while a junction zone contains $N_{de/jz} \sim 20$ dynamic electrodes \cite{zhangOptimizationImplementationSurfaceelectrode2022}. The algorithm for sorting this more realistic trap of $m \times n$ qubits using parallel dynamic control is illustrated in Fig.~\ref{fig:realistic_2d_trap_sort}. In an array of $N = m \times n$ qubits, each step can be executed by using $4 \times N_{de/gz} + 4 \times N_{de/jz} \sim 120$ dynamic DACs and $N$ bits of information. The algorithm has an approximate runtime of $2m + kn$ time steps. Similar techniques can be applied to handle longer chains, different numbers of qubits per junction, and specialized zones.
% \footnote{For example, note that row sort in this trap with four types of zones proceeds the same as it did in a 1D array with two types of zones, albeit with more external DACs. Therefore, more specialized zones can be added without fundamentally modifying the rearrangement algorithm.}.

\subsection{Performance}
\label{sec:reconfig_timing}

The total duration of qubit reconfiguration depends on the time it takes to execute a single step of parallel swaps. This, in turn, depends on the duration of other primitives, such as ion splitting/merging/rotation/linear shuttling (in case of a 1D swap in Fig.~\ref{fig:switchable_swap} b)) or junction shuttling (in case of a 2D swap in Fig.~\ref{fig:2d_array_swaps} a)). As an order of magnitude, we assume a qubit swap duration of $t_0 = 100 \ \upmu \text{s}$, which is representative of the capabilities of today's small-scale quantum computers \cite{bermudezAssessingProgressTrappedion2017, burtonTransportMultispeciesIon2022}. However, significantly faster swaps are feasible in the future \cite{steaneHowBuild3002006}.

Furthermore, in a 2D array, the reconfiguration time depends on the number of zones $N = m \times n$, as well as on the number of qubits per junction $k$. Finally, the reconfiguration time crucially depends on the target permutation, with longer-range connectivity requiring more reconfiguration steps.

As a pessimistic estimate, we consider the worst-case duration of arbitrary qubit routing. For the algorithm in Sec.~\ref{sec:realistic_2d_reconfig}, it is easy to verify that, for any given $(N, k)$, the reconfiguration time is minimized when $m = \sqrt{k N/2}$ and $n = \sqrt{2N/k}$. In that case, arbitrary reconfiguration of an $N$-qubit array requires approximately $\sqrt{8 k N}$ swap steps.

\begin{figure}[!ht]
\centering
{\includegraphics[width=0.45 \textwidth]{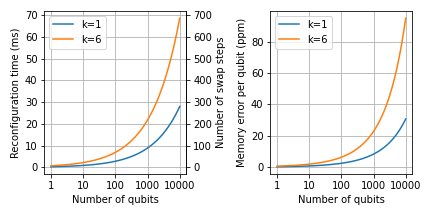}}
\caption{\label{fig:reconfig_time_and_error} 2D qubit array reconfiguration time and error. (Left) Approximate worst-case time and number of time-steps for reconfiguring a 2D array of qubits in the optimal configuration, assuming swap time of $t_0 = 100\ \upmu \text{s}$. (Right) Associated memory error, assuming the decoherence error model in \cite{sepiolProbingQubitMemory2019}.}
\end{figure}

With those assumptions, Fig.~\ref{fig:reconfig_time_and_error} (left) shows the maximum time necessary to perform arbitrary qubit routing in a realistic 2D array. We find that, with $k=6$, we achieve an arbitrary reconfiguration of a 1000-qubit array in $t_r = 22$ ms. To put that in more meaningful terms, we compute the memory error associated with qubit reconfiguration, assuming the error model measured for $^{43}$Ca$^+$ clock qubits in \cite{sepiolProbingQubitMemory2019}. The result is shown in Fig.~\ref{fig:reconfig_time_and_error} (right). We find that, for $k=6$ and $N=1000$, we can expect a memory error (per qubit per reconfiguration) of $2 \times 10^{-5}$, significantly below two-qubit gate errors in any quantum computing platform today.

This demonstrates that, despite limited freedom of operation, parallel dynamic control is compatible with high-fidelity reconfiguration at scale. At the same time, as each step requires $N$ bits of information to specify the switch settings, the overall required data flow rate between the controller and the chip is $N/t_0$, which evaluates to $10$ Mbit/s for a 1000-qubit chip. It is easy to send this amount of data over a single serial line, offering a major improvement over the standard approach. We emphasize that the numbers in this section describe routing times necessary for arbitrary all-to-all connectivity. In practice, judicious qubit mapping and algorithm choice can significantly decrease the overheads associated with qubit reconfiguration. Furthermore, specific reconfigurations can be executed much faster using less generic routing algorithms.

\subsection{Shim electrodes}
\label{sec:qubit_reconfiguration_shims}

The quasi-static shim electrodes play a passive role in qubit reconfiguration: they are charged to the desired values before reconfiguration begins, and are disconnected during ion transport. The primary function of shim electrodes is to minimize the stray field offsets between zones such that the same dynamic swap waveform leads to successful qubit swaps in different zones. Among all the transport primitives, ion splitting is typically most sensitive to the stray field setting, requiring axial stray field error of $E \lessapprox 5$ V/m \cite{negnevitskyFeedbackstabilisedQuantumStates2018}. However, larger stray-field errors can be tolerated as long they are radial. Finally, differences in stray fields can lead to heating and even ion loss during transport operations such as crystal rotations or junction transport. However, these can typically tolerate errors as large as $E \sim 100$ V/m \cite{burtonTransportMultispeciesIon2022}.

The second role of shim electrodes is to passively compensate for differences in dynamical electrode moments in different zones, for example, due to fabrication imperfections. While quasi-static shims cannot null out dynamical errors at every point in time, they can be used for fine-tuning the potentials along critical points in a transport trajectory, for example, to prevent frequency crossings during crystal rotations \cite{vanmourikCoherentRotationsQubits2020}.

Future experiments must verify the minimum number of necessary shim electrodes per zone. In subsequent sections, we conservatively assume each zone contains $N_{se/z} = 10$ shims\footnote{While placing $N_{se/z} \gg 10$ shims per zone increases the number of independent degrees of freedom, this has to be balanced against diminishing effectiveness of further sub-dividing the electrodes.}.
We anticipate that improvements in control design may allow for as few as $N_{se/z} \approx 3$ shims per zone, controlling 6 degrees of freedom (e.g. 3 field terms and 3 curvature terms) at one point in space during a two-qubit swap.

\section{Transport-assisted quantum gates}
\label{sec:transport_assisted_gates}
Consider an $N$-qubit trap, with a global spatially inhomogeneous qubit drive coupling to qubits in all zones. In order to perform transport-assisted quantum gates, we engineer a situation where, for every qubit $i$, the gate Rabi frequency $\Omega_{i}$ can be adjusted by adjusting the qubit position $r_i$. There are several different ways this can be implemented in hardware.

\begin{figure}[!ht]
\centering
{\includegraphics[width=0.45 \textwidth]{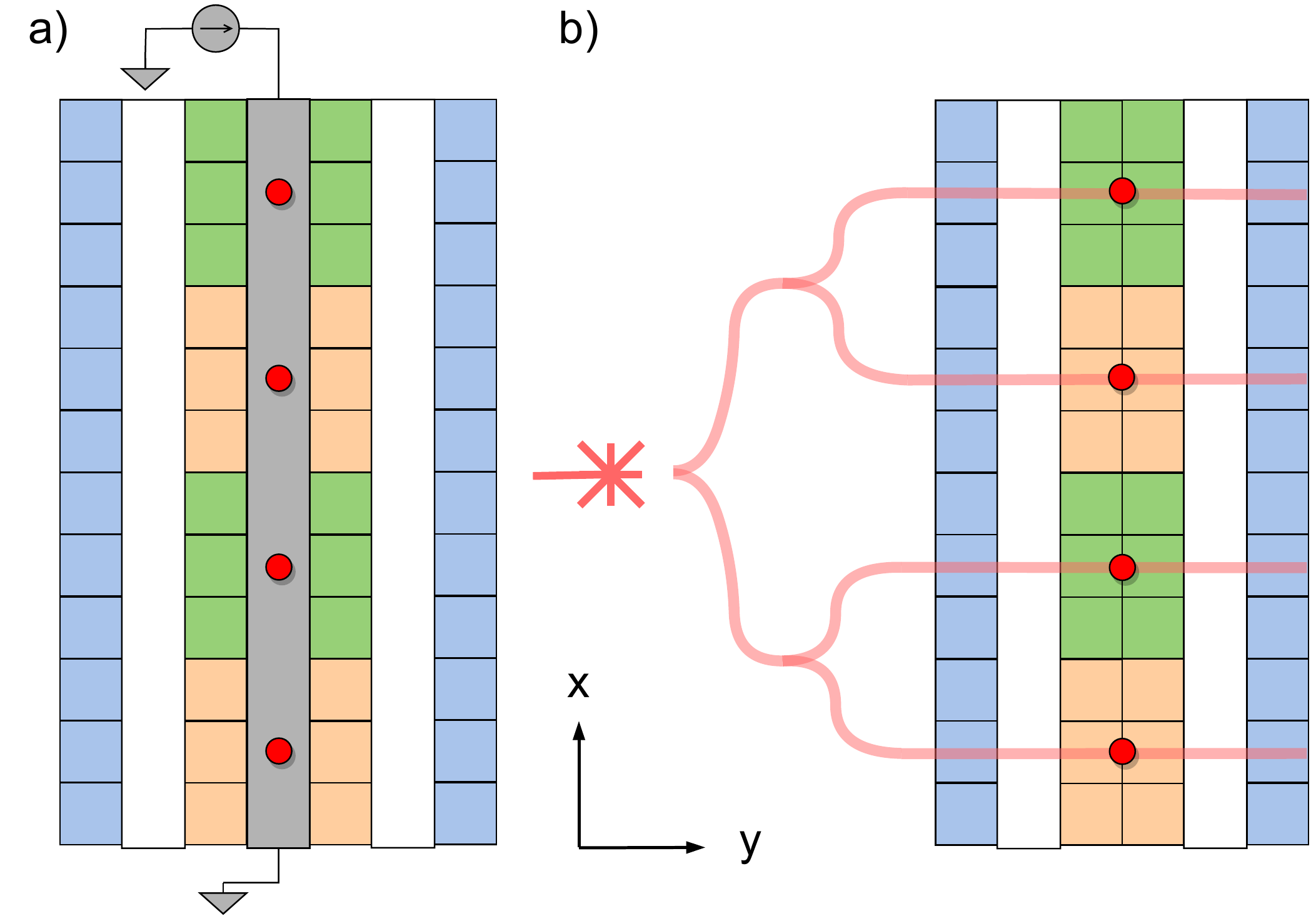}}
\caption{\label{fig:gate_microwave_laser} a) Schematic illustration of a linear ion trap supporting laser-free transport-assisted gates. Individual qubits (red) are held above a microwave conductor (grey) running along $x$ across all zones. When the current is applied, all qubits experience a position-dependent magnetic field, allowing for transport-assisted gates. b) Schematic illustration of a linear ion trap supporting laser-based transport-assisted gates. A laser beam is passively split into multiple channels, each delivered to a separate qubit in a separate zone. Once the laser is turned on, the qubits experience a position-dependent laser intensity, allowing for transport-assisted gates.}
\end{figure}

For example, one way to implement this scenario is by using integrated microwaves. A simplified case with a linear trap and a single conductor is shown in Fig.~\ref{fig:gate_microwave_laser} a). A trace carrying current $I$ along the trap axis $x$ creates a magnetic field $B \propto I$, near-resonant with the qubit frequency, in every zone in parallel. This leads to qubit coupling with Rabi frequency $\Omega(y) \propto I \times y$, where $y$ is the distance from the trap axis. Furthermore, it is possible to achieve $\Omega(y=0) \approx 0$, for example by exploiting polarization selection rules \cite{weberCryogenicIonTrap2022}, or using multiple parallel microwave lines \cite{ospelkausTrappedIonQuantumLogic2008, allcockMicrofabricatedIonTrap2013}. Thus, we obtain $\Omega(y) \approx \alpha I y$, where $\alpha$ is a constant. Therefore, single-qubit interactions can be switched off by placing the qubit at $y = 0$ and effectively modulated by moving the qubit along $y$ \cite{warringIndividualIonAddressingMicrowave2013, leuFastHighfidelityAddressed2023}. 

Another way to implement this scenario is using laser light, for example as illustrated in Fig.~\ref{fig:gate_microwave_laser} b). A laser beam, near-resonant with the qubit transition, is coupled into a trap-integrated waveguide \cite{mehtaIntegratedOpticalMultiion2020, niffeneggerIntegratedMultiwavelengthControl2020, ivoryIntegratedOpticalAddressing2021} and passively split between multiple paths of comparable intensity \cite{vasquezControlAtomicQuadrupole2023}. Inside each zone, a grating coupler is used to focus the light near the qubit location at $x = 0$ to a spot size $w$ \cite{mehtaPreciseDiffractionlimitedWaveguidetofreespace2017}, which leads to position-dependent Rabi frequency $\Omega(x) \propto \Omega_0 \exp(-x^2/w^2)$. Thus, single-qubit interactions can be modulated by moving the qubit around $x = \pm w/\sqrt{2}$, and switched off by placing the qubit at $x \gg w$. 

\subsection{Dynamic electrode parallelization}
\label{sec:transport_gates_parallel_dynamical}

\begin{figure}[!ht]
\centering
{\includegraphics[width=0.45 \textwidth]{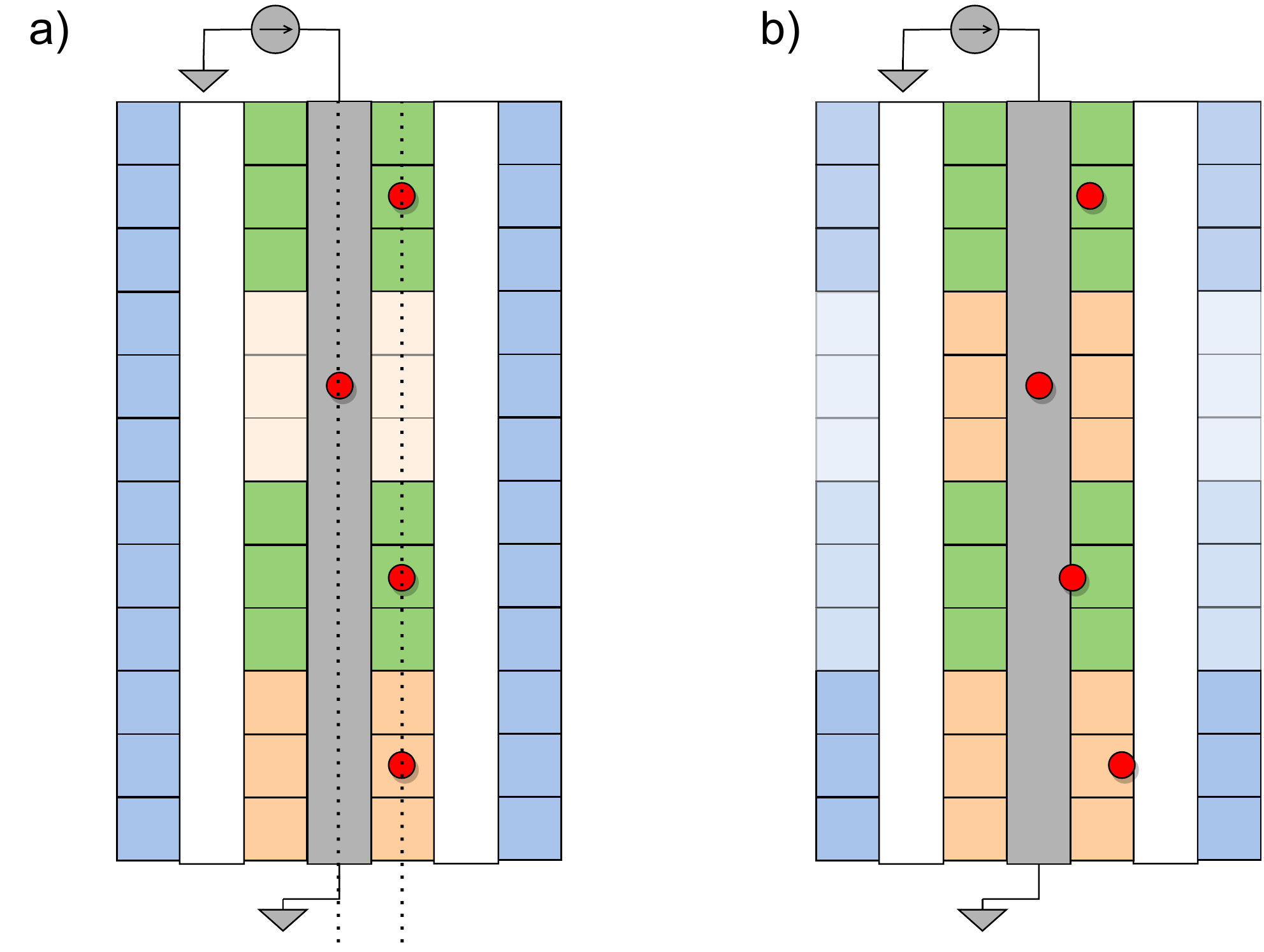}}
\caption{\label{fig:gate_parallel_multiplex} a) Illustration of transport-assisted laser-free single-qubit gates using a) dynamic electrode parallelization, and b) shim demultiplexing. In a), a switch select word $s$ selects which zones are active ($s=1$, solid color), and which are inactive ($s=0$, semi-transparent). Once the switches are set, a dynamical waveform moves the qubits in active zones to locations where they experience non-zero Rabi frequency $\Omega$, while qubits in inactive zones are moved to locations where $\Omega_{1q} \approx 0$. Afterward, a current pulse is applied to the central conductor, executing the target operation. Finally, the qubits are returned to their original locations. In b), the multiplexer is first turned on, setting shim electrodes in different zones to different values (shades of blue). This moves qubits in different locations to different zones, adjusting the Rabi frequency locally. Afterward, the multiplexer is turned on, and a current pulse is applied to the central conductor, executing the target operations.}
\end{figure}

We can leverage dynamic electrode parallelization for single-qubit gates, regardless of the implementation, as follows. We set up the dynamic waveforms such that, when $s_i=1$, the qubit in zone $i$ is moved to position $r_1$ with Rabi frequency $ \Omega$. On the other hand, if $s_i = 0$, the qubit is moved to position $r_0$ where $\Omega = 0$. An illustration of the scenario is shown in Fig.~\ref{fig:gate_parallel_multiplex} a). 

With the drive resonant with the qubit frequency, this generates a Hamiltonian $H_{\phi} = \sum s_i \ \Omega \ \sigma_{\phi}^{(i)}$, where $\sigma_{\phi} = \cos{\phi} \ \sigma_x + \sin{\phi} \ \sigma_y$, $\phi$ is the global drive phase, and $\Omega$ is the global drive strength. If the drive is detuned from the qubit transition frequencies by $\delta \gg \Omega$, we instead obtain a Hamiltonian $H_z = \sum_i s_i \sigma_{z} \ \Omega^2/(2 \delta)$ \cite{haeffnerQuantumComputingTrapped2008}.

Due to unavoidable experimental imperfections, it is impossible to ensure $\Omega = 0$ when $s_i = 0$. Thus qubits in zones with $s_i = 0$ will always experience some residual Rabi frequency $\Omega_{\varepsilon} = \varepsilon \times \Omega$. This, if uncorrected, results in $\sim \varepsilon^2$ infidelity per qubit per gate, which can be a challenge for transport-assisted gates. We mitigate this error using a two-step approach. First, we make use of any knowledge of $\varepsilon$ to coherently undo undesired rotations with subsequent operations. Second, we use composite pulse schemes to cancel out any residual but unknown systematic errors. Using the SK1 pulse sequence  \cite{merrillProgressCompensatingPulse2012}, for example, one can achieve single-qubit gate infidelities $< 10^{-5}$ with addressing imperfections as large as $\varepsilon \approx 0.1$, which is readily achievable in practice \cite{craikHighfidelitySpatialPolarization2017, wangHighfidelityTwoqubitGates2020, pogorelovCompactIonTrapQuantum2021}

In addition to single-qubit gates, dynamic electrode parallelization allows us to switch two-qubit interactions on- and off. When the global drive is turned on to generate a state-dependent force, we can write the effective entangling interaction between qubits $i$ and $i+1$ as $H = \Omega \ \sigma_{\phi}^{(i)} \sigma_{\phi}^{(i+1)}$. This interaction only occurs if the qubits are in the same potential well\footnote{$H$ only implements a global single-qubit rotation when applied to individual qubits, which can be corrected for in subsequent gates}. Therefore, we set up dynamical waveforms such that, if $(s_i, s_{i+1}) = (1,1)$, qubits $(i, i+1)$ are merged into the same well prior to the two-qubit gate, while if $(s_i, s_{i+1}) = (0,0)$, they remain in separate zones. Thus, we generate a Hamiltonian $H_2 = \sum_i s_{2i-1} \Omega \ \sigma_{\phi}^{(2i-1)} \sigma_{\phi}^{(2i)}$.

Hamiltonians $H_\phi$ and $H_z$ implement single-qubit gates on arbitrary subsets of qubits in parallel, while $H_2$ implements two-qubit gates on arbitrary subsets of qubits in parallel. Together, they can be used to implement a universal primitive gate set \cite{barencoElementaryGatesQuantum1995, nielsenQuantumComputationQuantum2010}, and thus universal quantum computation. Explicitly, a quantum circuit is decomposed onto $N_g$ different basis gates, and implemented by using cycles of $N_g$ time steps, each implementing one of $N_g$ gate layers.

\subsection{Shim demultiplexing}
\label{sec:transport_gates_shim_demux}
While dynamic electrode parallelization can be readily applied to circuits with a small number of basis gates $N_g$, it is prohibitively costly for circuits with $N_g \gg 1$. This creates a challenge in the NISQ regime, where the ability to implement a continuous variable-angle gate set is beneficial \cite{bullockArbitraryTwoqubitComputation2003, vatanOptimalQuantumCircuits2004, shendeMinimalUniversalTwoqubit2004}. Shim demultiplexing can be used to overcome this limitation of dynamic electrode parallelization by allowing continuous local control of $\Omega$. This leads to the ability to perform different single-qubit gates in parallel.

In this approach, instead of moving the qubits between two fixed positions $r_1$ and $r_0$, we adjust $r$ continuously using demultiplexed shim electrodes, as illustrated in Fig.~\ref{fig:gate_parallel_multiplex} b). After the shims are set, we turn off the demultiplexer and execute the operation by applying the global qubit drive. The shims are then set to new values before the next gate layer or transport layer.

Thus, shim demultiplexing allows us to implement a Hamiltonian $H_{\phi} = \sum_i \Omega_i \ \sigma_{\phi}^{(i)}$, where $\Omega_i$ is locally adjustable. This executes parallel rotations with locally adjustable angles across the processor using only a global drive. Combined with global two-qubit gates implemented as before, we obtain a powerful toolbox for NISQ algorithms and beyond.

There are several drawbacks associated with transport-assisted gates via shim demultiplexing. First, as electrode voltages have to be adjusted before every gate layer, demultiplexing slows down the effective clock rate of the quantum computer. Second, the lack of local phase control restricts the use of composite pulses to reduce systematic errors. However, local phase control can be restored using demultiplexed integrated IQ mixers, implementing the control scheme proposed in \cite{srinivasCoherentControlTrapped2022}. 

\section{Physical implementation}
\label{sec:implementation}
Ion trap fabrication and operation are fundamentally compatible with electronic integration and CMOS processes. However, as discussed in the introduction, care must be applied to issues of power dissipation, footprint, and data flow. Furthermore, low-complexity electronics are preferred given the relative immaturity of cryogenic CMOS design tools. Finally, care must be taken to avoid introducing additional error sources, such as RF pickup or stray fields.

In this section, we sketch out the structures that need to be integrated into the QPU to implement the WISE architecture. We discuss the design and main performance metrics of the dynamic electrode switch network (Sec.~\ref{sec:implementation_dynamic_parallelization}) and the shim demultiplexing network (Sec.~\ref{sec:implementation_shim_demultiplexing}). We discuss the main error sources, and in Sec.~\ref{sec:dac_comparison}, summarise why the physical implementation of the WISE architecture indeed alleviates all the main issues of integrated DACs.

\subsection{Dynamic electrode switch network}
\label{sec:implementation_dynamic_parallelization}

\subsubsection{Switch implementation}
\begin{figure}[!ht]
\centering
{\includegraphics[width=0.45 \textwidth]{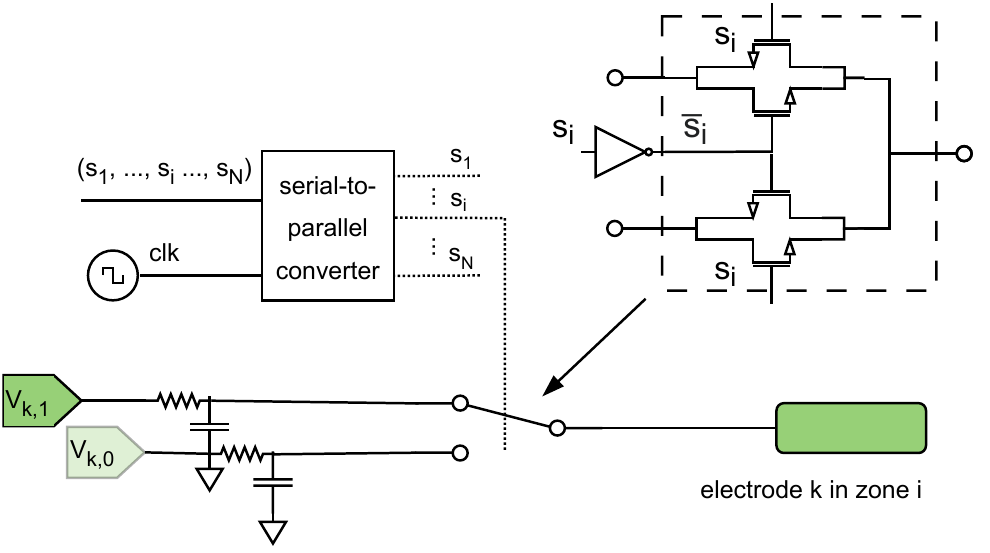}}
\caption{\label{fig:dynamic_switch_implementation} Basic implementation of a dynamic electrode switch network. The bit select word $s$ is loaded into a serial-to-parallel converter, which connects to a single-pole double-throw switch, selecting whether the electrode is active ($s=1$, solid) or inactive ($s=0$, transparent). The switch is implemented using two transmission gates, each implemented as a pair of transistors. 
}. 
\end{figure}

The basic building block of the dynamic electrode switch network is shown in Fig.~\ref{fig:dynamic_switch_implementation}. 
The select bit word $s = (s_1, \ldots, s_N)$ is loaded via a serial line into a parallel shift register. Afterwards, the entry $s_i$ is used to select whether the electrode $e = 1, 2, \ldots, N_{de/z}$ in zone $i$ is connected to DAC $V_{e,0}$ or $V_{e,1}$. The switch itself is built using two transmission gates, each implemented using an NMOS/PMOS transistor pair and an inverter. The DAC outputs are predominantly filtered off-chip.

\subsubsection{Switch network}
\begin{figure}[!ht]
\centering
{\includegraphics[width=0.45 \textwidth]{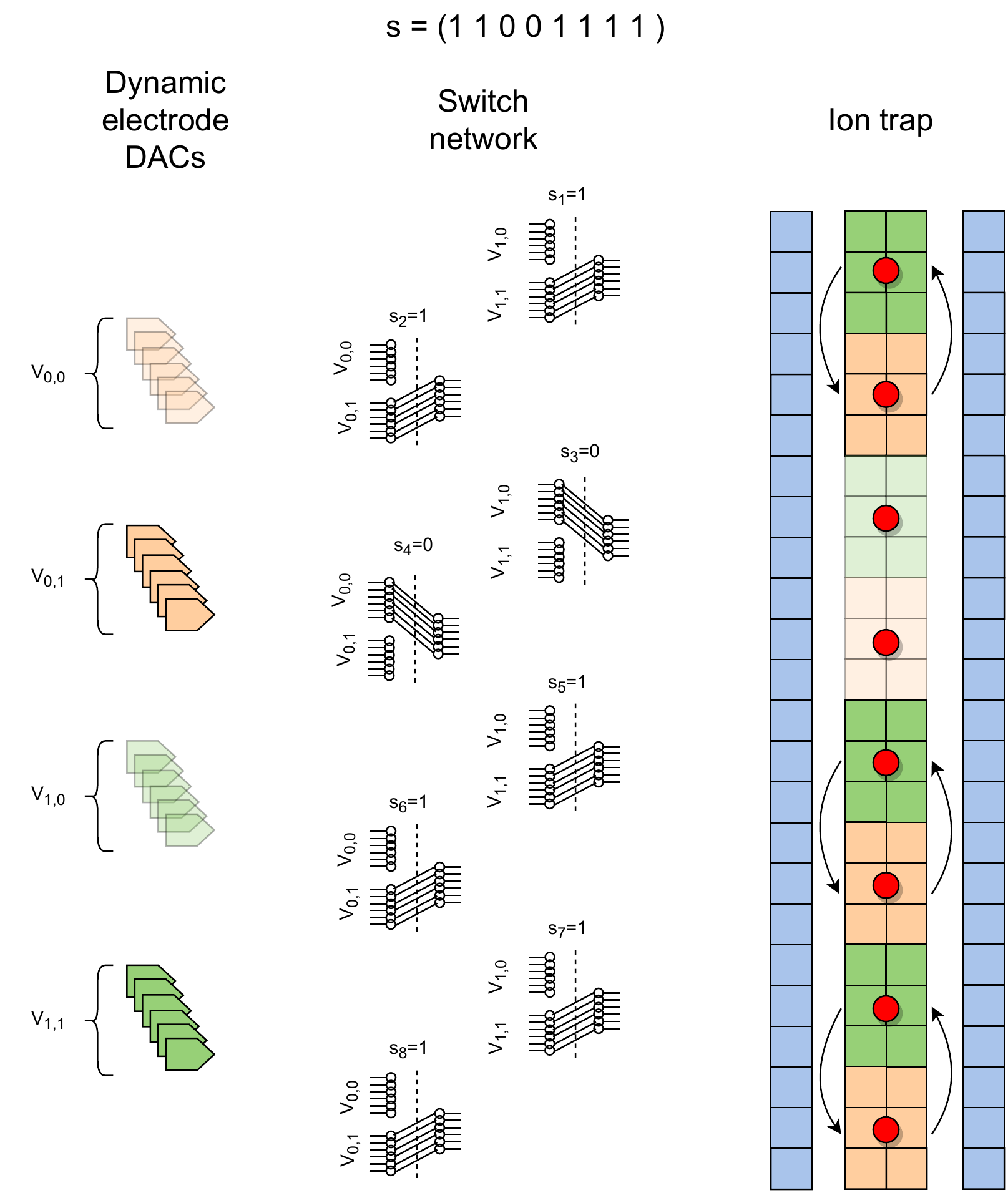}}
\caption{\label{fig:multi_switch_implementation} Implementation of parallel dynamic control. The linear 1D trap (right) is undergoing an odd swap. Each zone consists of dynamic electrodes, shown in green for odd zones, and in orange for even zones. Each dynamic electrode is connected through an individual single-pole double-throw switch to one of two DACs. All switches for zone $i$ are controlled by the same select bit $s_i$. Electrodes in odd zones are connected either to $V_{1,0}$ (if $s_i = 0$), or to $V_{1,1}$ (if $s_i = 1$). Likewise, electrodes in even zones are connected either to $V_{0,0}$ (if $s_i = 0$), or to $V_{0,1}$ (if $s_i = 1$). In the image above, the electrode color matches the color of the DAC it is currently connected to. In this (toy) example trap with $N_{de/z} = 6$ dynamic electrodes per zone, there are a total of $N_{dDAC} = 4 \times N_{de/z} = 24$ dynamic electrode DACs.}

Since each electrode is always connected to at least one DAC, there is no requirement to integrate capacitors, as off-chip capacitors suffice to provide a DC electrode RF ground. If necessary, residual RF pickup on dynamic electrodes can be reduced by optimizing transmission gate resistance.
 
\end{figure}

While in principle one switchable electrode per zone suffices to implement a switchable swap (Appendix \ref{sec:single_switch}) the most flexible wiring is illustrated in Fig.~\ref{fig:multi_switch_implementation}, where all $N_{de/z}$ dynamic electrodes in every one of $N$ zones are connected to switches. This allows for maximum flexibility in the waveform design, as every action can have a completely separate custom waveform set. As all the switches in zone $i$ are controlled by the same digital line $s_i$, the length of the bit select word $s$ is $N$.

\subsubsection{Operation timing}
The switch network is operated as follows. Consider a single operation step, either a swap (duration $t_0$) or a transport-assisted gate (duration $t_{tag}$). While the DACs are playing the required waveforms, the word $s$ of length $N$ bits describing the next transport step is loaded through a serial link into a parallel register. At the end of the transport step, the DAC voltages are set such that $V_{t,0} = V_{t,1}$, i.e. each electrode is at the same voltage regardless of the switch setting. Then, data from the parallel register is transferred onto the switches\footnote{Note that setting the switches does not affect electrode voltages, thus does not heat the ions.}, and the next transport step begins. Thus, assuming $t_{tag} > t_0$, as long as the interface supports a data rate of more than $N/t_0$, the switch network does not bottleneck the device performance. With $t_0 = 100\ \upmu \text{s}$ as in Sec.~\ref{sec:reconfig_timing}, we find that the required data flow rate is $10 \ N$ kbit/s. Thus, a single serial line with a bandwidth of $50$ Mbit/s suffices for $N \approx 5000$ qubits.

\subsection{Shim demultiplexing}
\label{sec:implementation_shim_demultiplexing}
\subsubsection{Demultiplexing network}
\begin{figure*}[!ht]
\centering
{\includegraphics[width=0.9 \textwidth]{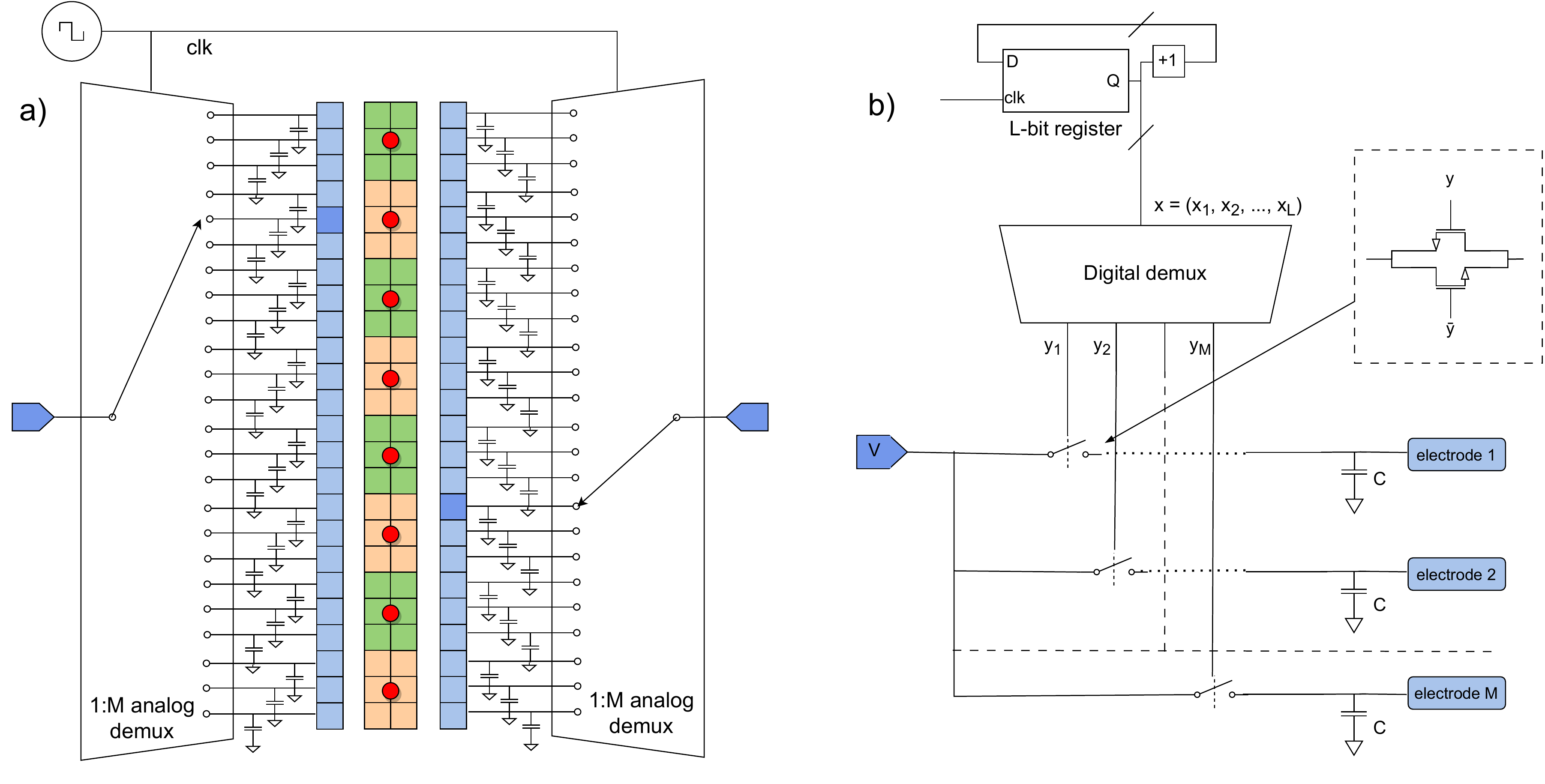}}
\caption{\label{fig:demultiplexer} a) Basic architecture for shim demultiplexing. Each shim DAC (blue) is connected to one of $M$ possible shim electrodes (blue) via an analog demultiplexer. An input clock controls all the demultiplexers on the chip, selecting whether they're turned on or off. Each shim electrode is connected to ground through an integrated capacitor. b) Details of analog demultiplexer implementation. An $L$-bit register (one for the whole chip) advances $x$ by 1 every clock cycle, The value of $x$ is sent through a digital demultiplexer and controls a transmission-gate switch, selecting which of the $M=2^L-1$ electrodes is connected to the DAC. When $x=0$, all the electrodes are disconnected, and the qubit reconfiguration / quantum gates can proceed.}
\end{figure*}

A high-level illustration of the shim demultiplexing network is shown in Fig.~\ref{fig:demultiplexer} a). The QPU consists of multiple analog $1:M$ demultiplexers that serve to connect the shim DAC to the shim electrodes one at a time. A single on/off signal clocks all the demultiplexers on the QPU. Each shim electrode is connected to an integrated capacitor after the demultiplexer.

Fig.~\ref{fig:demultiplexer} b) illustrates the demultiplexer circuit in more detail. On every clock cycle, on-chip digital logic increments an $L$-bit register $x$ by one, starting from $x=0$. The output $x = (x_1, x_2, \ldots, x_L)$ of the register is sent to a digital demultiplexer with $M = 2^L$ outputs. The circuit sets the output $y_x = 1$, while the rest of its outputs are set to 0. Demultiplexer outputs $y_x$ act as switch control signals, connecting the shim DAC to electrode $x$ iff $y_x = 1$. The individual switches are implemented as transmission gates as before.

After the switch is closed, we wait for time $t_{ec}$, allowing the electrode to charge, and then advance the clock by one cycle. The procedure is repeated in total $M$ times, charging $N_{se}$ shim electrodes in time $t_{sc} = t_{ec} \times M$ using $N_{se}/M$ DACs and a single clock line. When $x=0$, and all the electrodes are charged, we pause the clock signal and execute the subsequent operation layer with shim electrodes floating.

\subsubsection{Integrated capacitors}
\label{sec:capacitors}

Operation with floating electrodes requires integrated capacitors to provide an RF ground. To make the design practical, capacitances must be sufficiently small to avoid footprint bottlenecks highlighted for integrated DACs.

When disconnected, the voltage on each shim electrode oscillates with amplitude $V_{s,RF} = V_{RF} \times C_{s,RF}/C$, where $V_{RF}$ is the voltage on the RF electrode, and $C_{s,RF}$ is a parasitic capacitance between the RF and the shim electrode, and $C$ is the capacitance of the integrated capacitor. This leads to additional micromotion, especially if there are variations of $C_{s,RF}/C$ between electrodes.

Consider zones of area $A_z$, each containing $N_{se/z}$ shim electrodes, each connected to a capacitor of area $A_c$. As long as $N_{se/z} \times A_c < A_z$, the capacitors do not represent a significant bottleneck on the device size. Thin-film planar capacitors allow for $C/A_C \sim 3 \ \mathrm{fF}/\mu\mathrm{m}^2$~\cite{allcockHeatingRateElectrode2012}. Thus, a capacitor with dimensions $A_c = 100 \ \mu\mathrm{m} \times 100 \ \mu\mathrm{m}$ provides a capacitance of $C \sim 30$ pF. Assuming $A_z = 400 \ \mu\mathrm{m} \times 400 \ \mu\mathrm{m}$, we can connect $N_s \approx 16$ shims per zone to such capacitors without increasing  device footprint. Assuming $V_{RF} = 100$ V and $C_{s,RF} = 1$ fF \cite{doretControllingTrappingPotentials2012}, we expect an RF pickup of  $V_{s,RF} = 3$ mV. This should be sufficient for high-fidelity quantum operations, especially since the pickup can be very uniform with integrated capacitors. If necessary, RF pickup can be further reduced by using larger values of $C$, e.g. by increasing device footprint, increasing the number of capacitor layers, or by using vertical trench capacitors \cite{guiseBallgridArrayArchitecture2015, allcockHeatingRateElectrode2012}.

\subsubsection{Timing and I/O count}
\label{sec:electrode_charging}
The shim demultiplexing architecture allows us considerable freedom in trading off the shim charging speed, network complexity, performance, and the number of inputs. Increasing the multiplexing order $M$ decreases the number of shim DACs but increases the shim charging time $t_{sc}$, slowing down the effective clock speed of the quantum computer. While decreasing the electrode charging time $t_{ec}$ directly improves the clock speed, this must be traded off against the additional design and fabrication complexity of high-speed electronics. In addition, faster electrode charging can lead to additional motional excitation of ions, necessitating subsequent recooling. Finally, depending on the number of primitive gates $N_g$, single-qubit operations will either be executed using dynamic electrode parallelization (which doesn't require frequent shim recharging), or shim demultiplexing (which requires shim recharging between every gate layer).

As an example, consider a $N=1000$-qubit chip with $N_{se/z} = 10$ shims per zone, reconfiguration time $t_r = 22$ ms and $N_l = 10$ transport-assisted gate layers between reconfiguration steps, each requiring all shim electrodes to be recharged. Using a charging time of $t_{ec} = 3 \ \mu\mathrm{s}$ and $M=128$, the shim charging time is $t_{sc} = M \times t_{ec}$ = $384\ \upmu \text{s}$. Thus, shim charging takes $N_l \times t_{sc} / t_r \approx 20 \%$ of system time -- significant, but not a runtime bottleneck. At the same time, the total of $N_{se} = N_{se/z} \times N \sim 10,000$ shim electrodes will require $N_{se}/M \sim 80$ shim voltage inputs, easily achievable in today's systems. Finally, the charging time of $t_{ec} = 3 \ \mu\mathrm{s}$ is long enough to not introduce significant electronics design challenges, and to limit motional excitation of the ions.

% $M = 128$, we can thus charge $N_{se} = N_{sDAC} \times 128$ electrodes from $N_{sDAC}$ shim DACs in $t_{sc} = M \times t_{ec}$ = $380\ \upmu \text{s}$. In a $N=1000$ qubit chip as above, this increases the runtime by  $N_l \times t_{sc} / t_r \approx 20 \%$. At the same time, since $1/t_{ec} = 333$ kHz is below typical motional frequencies of small trapped-ion registers (1-10 MHz), undesired motional excitations should be limited. If necessary, dynamic electrodes can be used to move ions away from the active shim electrodes as they're being changed, decreasing the motional excitation further. Finally, the runtime can be decreased by only adjusting a subset of shims during gate operations.

\subsubsection{Voltage errors}

In our architecture, shim voltage errors can arise on top of errors from the DAC. The three main sources are charge injection, electrode discharging, and the photoelectric effect.

Charge injection refers to an additional voltage that appears once the switch is opened due to the redistribution of charge accumulated on the control transistors. This results in a fractional voltage shift of  $\sim C_t/C$, where $C_t$ is the transistor gate-source capacitance.
Assuming shim capacitance $C = 30$ pF, transistor capacitance $C_{t} = 30$ fF, and typical shim stray field $E_s = 200$ V/m \cite{guiseBallgridArrayArchitecture2015}, the charge injection effect is at the level of $E_s \times C_t/C \sim 0.2$ V/m. The effect is expected to be systematic, and thus compensatable by calibration. Furthermore, the magnitude of charge injection can be reduced by reducing $C_t$ through advanced fabrication processes.

Electrode discharging means electrode voltage $V$ decays over time once the switch is opened. The main loss channels are due to current flow through the capacitor and the transistor. We will assume the combined decay time constant of $\tau = 180$ s, based on results reported in \cite{xuOnchipIntegrationSi2020} for cryogenic transistors and thin-film alumina capacitors of comparable values.

To evaluate the effect of electrode discharging on quantum operations, consider once again a $N=1000$ qubit chip with reconfiguration time $t_r = 22$ ms (as in Sec.~\ref{sec:reconfig_timing}) and two-qubit gate time $t_{2q} = 100 \ \upmu \text{s}$. Assuming once again $E_s = 200$ V/m, we find a shim field drift of $E_s \times (1-e^{-t_{r}/\tau}) \approx 24$ mV/m during reconfiguration, sufficient for low-heating transport. The most stringent requirement on electrode voltage stability is that stray field drifts can lead to gate mode frequency drifts, causing two-qubit gate errors. During a two-qubit gate, we expect a stray field drift $E_s \times (1-e^{-t_{2q}/\tau}) \approx 0.1$ mV/m.  Assuming gate mode anharmonicity of $df/dE = 1$ kHz per V/m \cite{homeNormalModesTrapped2011}, this corresponds to motional frequency drift of $df = 0.1$ Hz, resulting in a negligible two-qubit gate error of $\sim df^2 \times t_{2q}^2 \sim 10^{-10}$ \cite{malinowskiUnitaryDissipativeTrappedIon2021}. Thus, we expect that electrode discharging will not bottleneck either qubit reconfiguration or quantum gate fidelity\footnote{Also note that it is in principle possible to keep track of these decaying fields and apply appropriate corrections to the drive fields}.

Due to the photoelectric effect, the voltage held on the shim electrode -- and hence the corresponding stray field -- is perturbed by laser light. This effect is most pronounced for short-wavelength light (blue and ultraviolet), but can be meaningful for longer-wavelength radiation as well \cite{harlanderTrappedionProbingLightinduced2010}. The impact of the photoelectric effect can be minimized by carefully managing laser scatter and electrode contamination, as well as by using laser-free gates instead of laser-based gates.

\subsection{Comparison to integrated DACs}
\label{sec:dac_comparison}
In a nutshell, our solution replaces the problem of integrating filtered DACs with the problem of integrated switches and small capacitors. Thus, it is only a reasonable solution if the latter issue is easier than the former. Fortunately, as highlighted earlier in the section, our architecture alleviates all the main concerns of integrated DACs:
\begin{enumerate}
    \item Power dissipation. Unlike a DAC, a transmission-gate switch results in negligible additional power consumption. 
    \item Data bandwidth. Unlike a DAC, a transmission-gate switch requires only a single bit of information per operation. Thus, the data streaming rate is reduced to $1/t_0 = 10$ kbit/s per zone, allowing for a large number of switches with a single serial line.
    \item Footprint. Unlike a DAC, a transmission-gate switch requires a footprint only a few times larger than a single transistor. The transistor size, in turn, depends on the required voltage level, as well as the details of the fabrication process. Assuming a transmission gate footprint of $A_{tg} \approx 50 \ \upmu \text{m} \times 50 \ \upmu \text{m}$, realistic for $\pm 10$ V logic levels \cite{stuartIntegratedTechnologiesControl2021}, we find that the switching networks introduce little-to-no overheads on the device footprint (Appendix \ref{sec:app_table}). Furthermore, all large DAC filter capacitors (with typical values of $\sim 1$ nF) are placed off-chip in our architecture.
    While WISE architecture requires integrated capacitors to eliminate RF pickup on shim electrodes, those can be small ($C \sim 30$ pF), and their footprint manageable ($A_C \approx 100 \ \upmu \text{m} \times 100 \ \upmu \text{m}$).
\end{enumerate}
All in all, integrating switches at high density is considerably more practical than integrating DACs at high density.

\section{Wiring a 1000-qubit chip}
\label{sec:wiring_1000_qubits}

\begin{figure*}[!ht]
\centering
{\includegraphics[width= \textwidth]{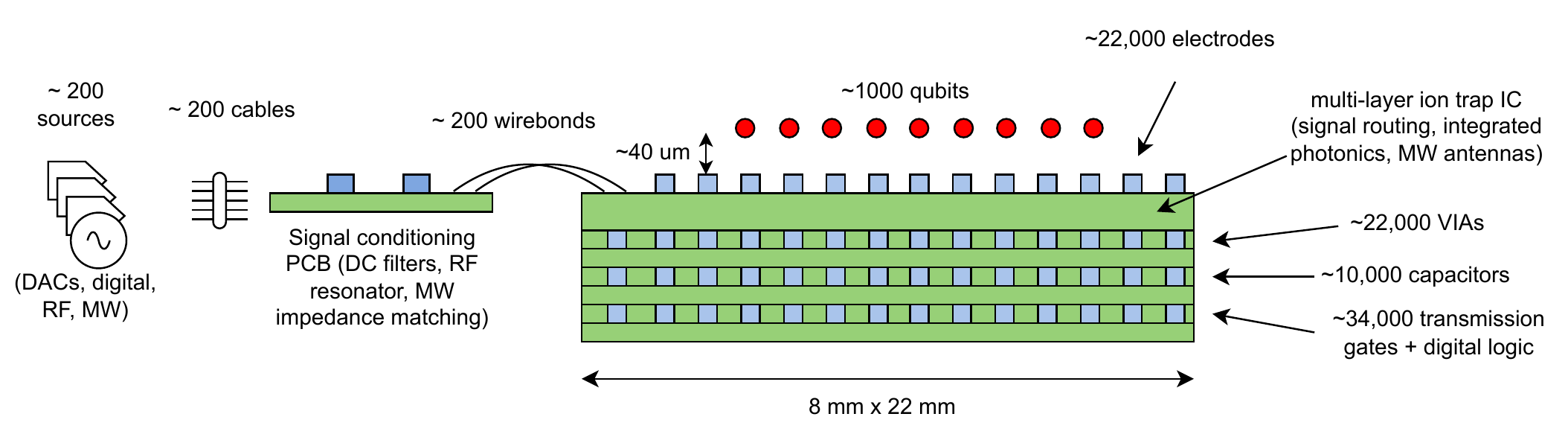}}
\caption{\label{fig:wiring_1000_qubits} Illustration of possible electrical wiring of a 1000-qubit chip. The QPU (right) combines an ion trap IC with integrated capacitors and switches, and requires a footprint of $8 \text{mm} \times 22 \text{mm}$ (excluding interconnects).  The QPU is controlled using $\sim 200$ electrical inputs, delivered from $\sim 200$ off-chip sources via wire bonds.}
\end{figure*}

Let us now put all the numbers from the previous sections together to discuss the wiring of a 1000-qubit trapped ion quantum computer. Table \ref{tab:sys_params} in Appendix \ref{sec:app_table} shows the main system parameters, collected from the previous sections. In this section, we summarise the most important quantities. The result is illustrated in Fig.~\ref{fig:wiring_1000_qubits}.

Our device is a 2D array of $m \times n = 54 \times 19 = 1026$ zones, consisting of $N_{gz} = 855$ gate zones and $N_{jz} = 171$ junction zones. The device holds $N = 1026$ qubits ($k = 6$ qubits per junction) at a height of $h = 40$ um above the surface, in a region of $x \times y = 22 \times 8$ mm (total area $A_t = 164 \ \text{mm}^2$), using $N_{de} = 11,970$ dynamic electrodes and $N_{se} = 10,260$ shim electrodes. Each shim electrode is connected to one of the $C = 30$ pF thin-film planar capacitors, which occupy a total area of $A_{t,c} = 103 \ \text{mm}^2$ on a buried layer. Another buried layer contains the switching network -- a total of $N_{tg} = 34200$ transmission gates, at an area of $86 \ \text{mm}^2$ -- as well as the necessary digital logic.

After the switching network, the dynamic electrodes are connected to $N_{dDAC} = 120$ off-chip DACs. The shim electrodes are controlled by $N_{sDAC} = 81$ off-chip DACs via $1:128$ on-chip analog demultiplexers. In addition, $\sim 10$ electrical connections are required to deliver the digital control signals (with data streaming rate of $50$ Mbit/s), power the digital logic, as well as to provide global trapping and qubit drive fields. In total, we use $\sim 210$ electrical connections between the control electronics and the QPU, comparable to present-day ion trap setups. Additional interconnects will be used for optical wiring (for laser delivery and readout), but the discussion of these is out of the scope of this work.

Quantum operations are performed as interleaved layers of qubit reconfiguration and transport-assisted gates, with arbitrary-angle single-qubit rotations implemented by shim demultiplexing. 
The reconfiguration takes between $t_0 = 100 \ \upmu \text{s}$ (for simple nearest-neighbour routing) and $t_r = 22$ ms (for arbitrary large-scale routing). Quantum gates are enabled by global qubit drives, which implement single-qubit gates in $t_{1q} = 1 \ \upmu \text{s}$ and two-qubit gates in $t_{2q} = 100 \ \upmu \text{s}$. However, the duration of each
transport-assisted gate layer is dominated by the shim
charging time of $t_{sc} = 384 \ \upmu \text{s}$. Depending on the operation, the overall circuit execution speed is therefore between $1/(t_{1q}+t_{sc}) = 2600$ circuit layers per second (for sequential single-qubit gates in identical qubit configuration) and $1/(t_r+t_{2q}+t_{sc}) = 44$ circuit layers per second (for sequential two-qubit gates in maximally different qubit configurations).

\section{Summary}
We have shown how integrating simple switching electronics into the QPU allows for a 1000-qubit trapped ion quantum computer to be operated with only $\sim 200$ DACs. The QPU allows for transport-assisted gates, with low errors and effective all-to-all connectivity.

Today's trapped-ion systems already routinely use $\sim 200$ DACs per QPU \cite{decaroliDesignFabricationCharacterization2021}. Furthermore, today's ion traps are routinely made on silicon substrates \cite{choReviewSiliconMicrofabricated2015}, with multiple experiments demonstrating compatibility with substrate-integrated passive \cite{zhaoTSVintegratedSurfaceElectrode2021} and active \cite{reensHighFidelityIonState2022, setzerFluorescenceDetectionTrapped2021} electronics, as well as CMOS processes \cite{mehtaIonTrapsFabricated2014, auchterIndustriallyMicrofabricatedIon2022}. Therefore, the proposed design shows that we can increase the scale of trapped-ion quantum computers from today's $N \sim 10 - 30$ qubits to $N \sim 1000$ qubits without fundamental roadblocks in electronic wiring. This opens the path to useful NISQ-scale quantum computation with trapped ions.

\subsection{Challenges and open questions}

First and foremost, we must verify the feasibility of performing the same transport operations in different trap zones using fixed dynamical waveforms and local shims. This may require optimizing chip fabrication to minimize zone-to-zone variations of electrode moments and to shield stray charges. Second, processor zones and waveforms must be carefully designed to minimize zone-to-zone crosstalk, such that a transport operation in one zone succeeds regardless of the waveform being applied to other zones. If necessary, spacing between zones can be increased without increasing the number of dynamic DACs count using ``bucket-brigade" or ``conveyor-belt" type shuttling \cite{sangsterBucketbrigadeElectronicsNew1969, alonsoQuantumControlMotional2013, millsShuttlingSingleCharge2019, holzphilipIonlatticeQuantumProcessors2019, seidlerConveyormodeSingleelectronShuttling2022}. Third, the RF pickup on the dynamic electrodes caused by finite switch impedance may affect the transport operations in undesirable ways, necessitating the development of low-loss switches or the integration of additional capacitors.

The most important verification of shim demultiplexing is to confirm the stability of voltage offsets caused by charge injection. If problematic, on-chip switches can be miniaturized to reduce transistor capacitance, hence the stored charge. This can, however, lead to increased current leakage, in turn shortening the electrode discharging time. The second assumption that requires verification is that high-fidelity quantum gates can be performed with floating shim electrodes. Finally, experiments should establish realistic timescales for electrode charging, in turn guiding the decision of how many shim electrodes can be connected to a single DAC.

Future experiments must also verify the feasibility of high-fidelity transport-assisted gates. Despite proof-of-principle demonstrations, all highest-fidelity gates in trapped-ion systems were thus far executed with externally modulated localized drives \cite{hartyHighfidelityPreparationGates2014, srinivasHighfidelityLaserfreeUniversal2021, clarkHighFidelityBellStatePreparation2021}. Further work is needed to demonstrate that precise control over local potentials can be used to overcome zone-to-zone variations, enabling high-fidelity parallel gates. Furthermore, fast and robust calibration procedures must be developed to stabilize local stray-field drifts in larger-scale systems \cite{nadlingerMicromotionMinimisationSynchronous2021}.

\subsection{Larger-scale systems}

While this paper focused on wiring a 1000-qubit chip, it is interesting to ask if WISE can be applied to larger-scale systems. For example, could a million-qubit chip be wired in the same fashion? 

On the face of it, dynamic electrode parallelization can be used to control arbitrarily many qubits from a fixed number of DACs. In practice, this will be limited, for example, by parasitic capacitances of the switches and electrodes and the resulting DAC load swings. Another relevant effect is that the RF pickup on $N_{de/dDAC}$ dynamic electrodes connected to one DAC increases in proportion to $N_{de/dDAC}$, affecting the design of DAC filters and, in turn, transport speed.

On the other hand, shim demultiplexing always requires the number of shim DACs and QPU I/O lines to grow in proportion to the qubit number ($N_{sDAC} \sim N/10$ in the main text). Scaling the method to $N \gg 1000$ requires decreasing the proportionality factor, which can be achieved by a) increasing the multiplexing order above $M=128$ shims per DAC, and b) decreasing the number of shim electrodes per zone to $N_{se/z} < 10$. While the former increases total shim charging time $t_{sc}$, this is not an issue if the gates are implemented using dynamic electrode multiplexing and if the shim discharging timescales are long. If necessary, $t_{sc}$ can be reduced by decreasing the charging time per electrode $t_{ec}$, or by sequentially charging additional on-chip capacitors during quantum operations, and then connecting them in parallel to the shim electrodes. 

All in all, while WISE can be applied to systems with $N > 1000$ qubits, the architecture has its limitations which will bound the maximum number of qubits it can efficiently support to some number $N_{max}$. Regardless of its precise value (which is difficult to estimate), an $N > N_{max}$-qubit QPU can be then constructed as repeating units of $N_{max}$ qubits. While DAC and control system integration become necessary for large enough $N$, WISE architecture can nonetheless be employed to significantly reduce the number of DACs per qubit compared to the standard approach, making the power dissipation and footprint issues much more tangible.

To summarize, the WISE architecture opens the door to building trapped-ion quantum computers which are 1–2 orders of magnitude larger than today’s systems. Looking ahead to even larger devices, there are a number of architectural decisions that will significantly impact system wiring. For example, can we reduce the system resource cost if we specialize the WISE architecture for implementing a specific fault-tolerant quantum error-correction code, and if so, which quantum error correction code is optimal \cite{cohenLowoverheadFaulttolerantQuantum2022, tremblayConstantoverheadQuantumError2022}? What is the optimal chip size and connectivity for encoding logical qubits \cite{metodiQuantumLogicArray2005a, monroeLargescaleModularQuantumcomputer2014, nigmatullinMinimallyComplexIon2016}? Does the code benefit from long-range connectivity, and if so, of what kind \cite{stephensonHighRateHighFidelityEntanglement2020, gaoOptimisationScalableIonCavity2023, krutyanskiyEntanglementTrappedionQubits2023}? Regardless of the exact large-scale design, we believe that the methods behind WISE architecture – after thorough refinement in NISQ-scale devices – can serve as its backbone.

We are grateful to the entire team at Oxford Ionics for their support and helpful discussions.
\bibliography{mylib}

\appendix

\section {1000-qubit system parameter table}
\label{sec:app_table}
A detailed list of parameters of a 1000-qubit system and their derivation is shown in Tab.~\ref{tab:sys_params}.

\label{sec:param_tab}

\begin{table*}
\caption{\label{tab:sys_params} System parameter table for a 1000-qubit chip}
\begin{ruledtabular}
\begin{tabular}{lllr}
Quantity &
Symbol &
Formula  &
Value \\
\colrule
No. of qubits = No. of zones & $N$ & & 1026 \\
No. of qubits per junction & $k$ & & 6 \\
Zone count & $m \times n $ & $\approx \sqrt{k N/2} \times \sqrt{2 N/k}$ & $54 \times 19$ \\
No. of junctions zones & $N_{jz}$ & $n \times m/k$ & 171 \\
No. of gate zones & $N_{gz}$ & $N - N_{jz}$ & 855 \\
No. of dynamic electrodes per gate zone & $N_{de/gz}$ & & 10 \\
No. of dynamic electrodes per junction zone & $N_{de/jz}$ & & 20 \\
No. of shim electrodes per zone & $N_{se/z}$ & & 10 \\
No. of dynamic electrodes & $N_{de}$ & $N_{de/gz} \times N_{gz} +  N_{je/gz} \times N_{jz}$ & 11,970 \\
No. of shim electrodes & $N_{se}$ & $N_{se/z} \times N $ & 10,260 \\
No. of electrodes & $N_{e}$ & $N_{de} + N_{se} $ & 22, 230\\
\hline 
Ion height & $h$ & & $40 \ \upmu \text{m}$ \\
Average zone size & $x_z \times y_z$ & $10h \times 10h$ & $400 \ \upmu \text{m} \times 400 \ \upmu \text{m}$ \\
Chip size (active region) & $x \times y$ & $(m \times x_z) \times (n \times y_z) $ & $22 \ \text{mm} \times 8 \ \text{mm}$\\
Total area (active region) & $A_t$ & $x \times y$ & $164 \ \text{mm}^2$\\
\hline 
On-chip capacitance density & $C/A_C$ & & 3 $\text{fF}/\upmu \text{m}^2$ \\
On-chip capacitor size & $A_C$ & & $100 \ \upmu \text{m} \times 100 \ \upmu \text{m}$ \\
Total area (capacitors)  & $A{t,c} $ &  $A_C \times N_{se}$ & $103 \ \text{mm}^2$ \\
Shim capacitance to GND (when floating) & $C$ & $C/A_C \times A_C $ & $30$ pF 
\\
RF voltage & $V_{RF}$ &  & 100 V \\
Shim-to-RF capacitance & $C_{s,RF}$ &  & 1 fF \\
RF voltage on shim electrodes & $V_{s,RF}$ & $V_{RF} \times C_{s,RF} / C $ & 3 mV \\
\hline
No. of dynamic switch settings & $N_{\text{set}}$ & 0 or 1, i.e. stay or swap & 2 \\
No. of dynamic DACs & $N_{\text{dDAC}}$ & $2 \times N_{\text{set}} \times (N_{de/gz} + N_{de/jz})$ & 120 \\
Multiplexing order & $M$ &  & 128 \\ 
No. of shim DACs & $N_{\text{sDAC}} $ & $N_{se}/ (M-1)$ & 81 \\
Length of switch select word & $N_{sw}$ & $\log_2(N_{\text{set}} ^ N)$ & 1026 bits \\
Serial link data rate & $dN_{s}/dt$ & & 50 Mbit/s \\
Switch select time & $t_{ss} $ & $ N_{sw} / dN_{s}/dt$  & $21 \ \upmu \text{s}$ \\
No. of trans. gates per dynamic electrode & $N_{tg/de}$ &  & 2 \\
No. of trans. gates per shim electrode & $N_{tg/se}$ & & 1 \\
No. of trans. gates & $N_{tg}$ & $N_{tg/de} \times N_{de} + N_{tg/se} \times N_{se}$ & 34200 \\
Trans. gate size & $A_{tg}$ & & $50 \ \upmu \text{m} \times 50 \ \upmu \text{m}$ \\
Total area (trans. gates) & $A_{t,tg} $ & $N_{tg} \times A_{tg}$ & $86 \ \text{mm}^2$ \\
\hline
Charging time per shim electrode & $t_{ec}$ &  & $3 \ \upmu \text{s}$\\
Total shim charging time & $t_{sc}$ & $t_{ec} \times M$ & $384 \ \upmu \text{s}$\\
Qubit swap time & $t_{0}$ &  & $100 \ \upmu \text{s}$\\
Maximum qubit reconfiguration time & $t_{r}$ & $2\times m + k \times n$ & 22 ms\\
Transistor capacitance & $C_t$ &  & 30 fF \\
Typical shim field & $E_s$ &  & 200 V/m \\
Charge injection field systematic & $dE_{c}$ & $\sim E_s \times C_t/C$ & $\sim 0.2 $V/m \\
Shim  discharging time constant & $\tau$ & & 180 s \\
Shim field drift during reconfiguration & & $E_s \times (1-e^{-t_{r}/\tau})$ & 25 mV/m \\
Shim field drift during two-qubit gate pulse & & $E_s \times (1-e^{-t_{2q}/\tau})$ & 0.1 mV/m \\
\hline
1-qubit gate pulse time & $t_{1q}$ &  & $1 \ \upmu \text{s}$ \\
2-qubit gate pulse time & $t_{2q}$ &  & $100 \ \upmu \text{s}$ \\
Transport-assisted 1-qubit gate time & $t_{tag, 1q}$ & $t_{sc} + t_{1q}$ & $385 \ \upmu \text{s}$ \\
Transport-assisted 2-qubit gate time & $t_{tag, 2q}$ & $t_{sc} + t_{2q}$ & $484 \ \upmu \text{s}$ \\
Maximum system speed & & $1/t_{tag, 1q}$ & 2600 layers per second \\
Minimum system speed & & $1/(t_{tag, 2q}+t_r)$ & 40 layers per second

\end{tabular}
\end{ruledtabular}
\end{table*}

\section {Parallel dynamic control with one switch per zone}
\label{sec:single_switch}
The implementation presented in the main text assumes every dynamic electrode is connected to a separate switch. This is a conservative assumption, and it should be in fact possible to only connect some dynamic electrodes to switches while leaving the remainder of dynamic electrodes permanently co-wired. In this section, we present an example implementation of that idea, where every zone features only one switchable electrode.

\begin{figure}[!ht]
\centering
{\includegraphics[width=0.3 \textwidth]{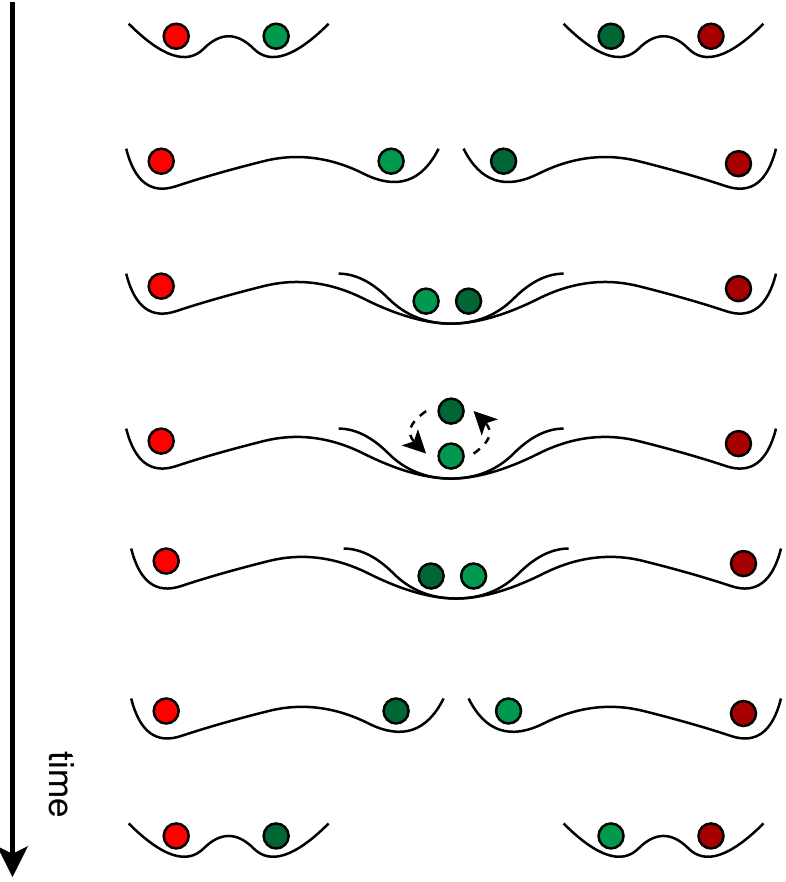}}
\caption{\label{fig:double_1d_swap_waveform} Extended 1D swap waveform. If there are four qubits (two green, and two red) at $t=0$, the green qubits undergo a swap, while the red qubits remain stationary. Alternatively, we can think of it as a two-qubit swap waveform: if there are two qubits in green locations at $t=0$, they undergo a swap, while if the qubits are initially at red locations, they remain stationary.}
\end{figure}

Consider an extension of the 1D swap waveform between zones $(1,2)$ shown in Fig.~\ref{fig:double_1d_swap_waveform}. In it, green qubits undergo a swap, but red qubits remain stationary. An alternative way to use this waveform is as follows. Suppose there are only two qubits - one in zone 1, and one in zone 2 - and a switchable ``push field" that can be used to push qubits in each zone either left or right. Specifically, if $s_i = 1$, the qubit in zone $i$ is pushed into the green location, while if $s_i = 0$, it is pushed into the red location. After the switch is set, the waveform is executed. Thus, if $(s_1,s_2) = (1,1)$, the qubits undergo a swap, while if $(s_1, s_2) = (0,0)$, they remain in the original order.

\begin{figure}[!ht]
\centering
{\includegraphics[width=0.45 \textwidth]{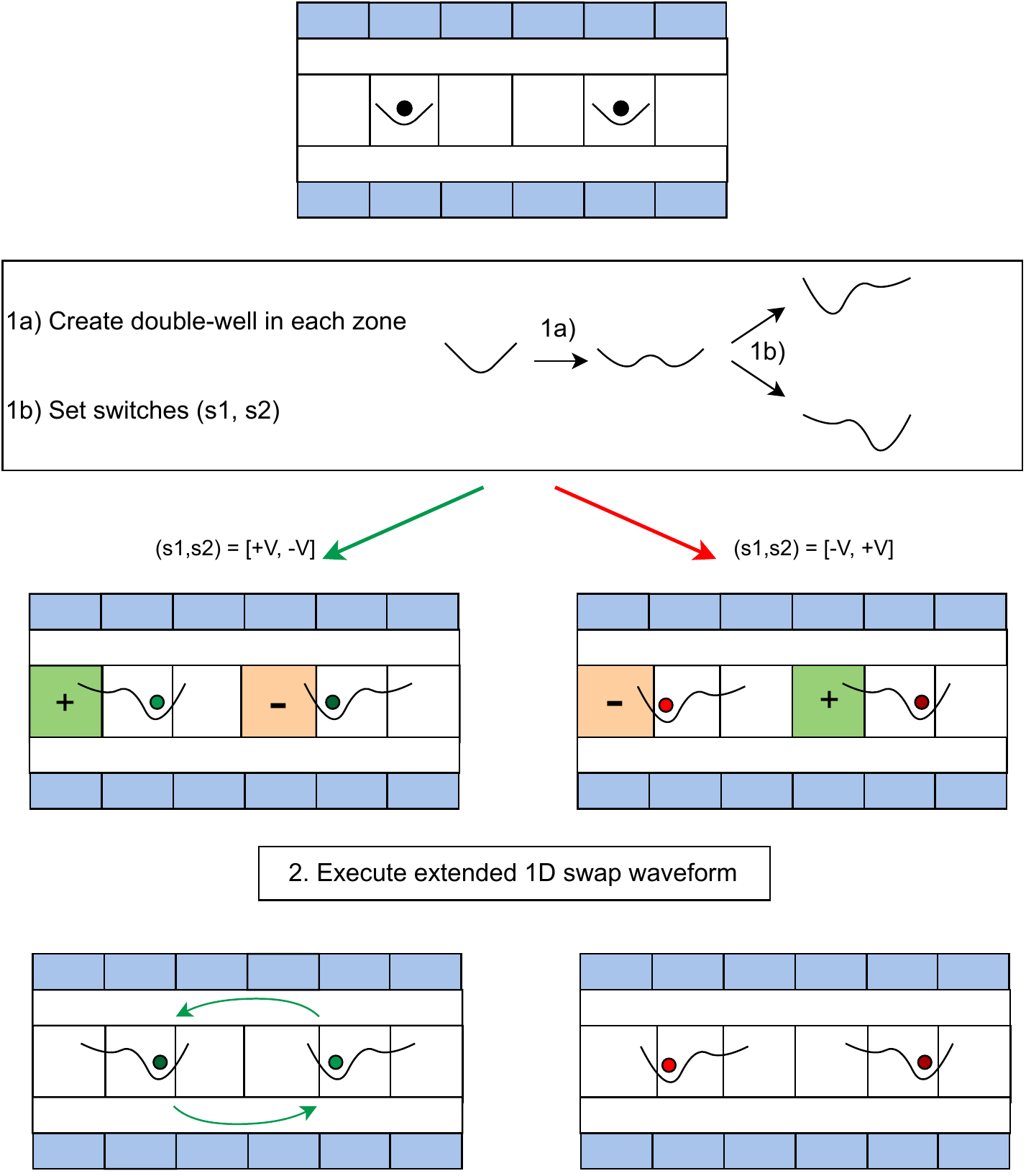}}
\caption{\label{fig:single_switch} Single-switch implementation of parallel digital control. Initially, two qubits (black) are trapped in two separate zones using two separate waveforms. The initial wells are evolved into double wells in every zone (1a), while local switches $s_i$ are used to select if the qubit in zone $i$ is pushed into the left or the right well (1b), thus selecting if the qubits will undergo a swap once the extended 1D waveform is applied. Steps 1a) and 1b) can be executed simultaneously or sequentially, in any order.}
\end{figure}

This swap waveform can be physically implemented as illustrated in Fig.~\ref{fig:single_switch}. We use dynamic electrodes co-wired to the same DACs to create a double-well potential (first step of the extended transport waveform) in each zone. At the same time, for each zone, we can select whether the ion is pushed into the red or green well at $t=0$ by applying either positive voltage $+V$ or negative voltage $-V$ to one electrode, as illustrated in Fig.~\ref{fig:single_switch}. Afterward, the dynamic DACs play the remainder of the extended transport waveform. This sequence allows us to locally and digitally select if the ion undergoes a swap or not. Thus, one switch per zone suffices to implement parallel dynamic control. The same method can be used to condition the execution of a 2D swap waveform.

In the basic implementation, the $\pm V$ rails connect to separate DACs and are initially set to the same voltage $V_0$. Once each electrode is connected to one rail, the voltage on the $+V$ rail is increased to $V_0+V$, and the voltage on the $-V$ rail is reduced to $V_0-V$. This ramping can be done smoothly to avoid undesired motional excitation, with DAC outputs filtered off-chip. After all electrodes are set to either $V_0+V$ or $V_0 - V$, the swap waveform is played, and the rails return to $V_0$. This implements one cycle of parallel swaps.

Alternatively, parallel switching can be implemented by purely digital means. In this method, the $\pm V$ rails are permanently set to $V_0 \pm V$. This can considerably simplify circuit design, and make it compatible with standard digital CMOS. However, a digital implementation may require capacitors after the switches to smooth out the switching impulses. Alternatively, qubits can be displaced with minimum motional excitation and without filtering by a sequence of switching events using bang-bang methods \cite{alonsoQuantumControlMotional2013,alonsoGenerationLargeCoherent2016}.

\end{document}